\documentclass[hidelinks,a4paper,10pt,onecolumn, switch]{article}
\pdfoutput=1

\usepackage{fancyhdr}
\fancypagestyle{plain}{%
	\fancyhf{}
	\fancyfoot[R]{\thepage}
	\fancyfoot[L]{\small \textit{Preprint}\\ \texttt{arXiv}}
}

\usepackage[affil-it]{authblk}
\makeatletter
\def\@maketitle{%
	\newpage
	\null
	\vskip 2em%
	\begin{center}%
		\let \footnote \thanks
		{\Large\bfseries \@title \par}%
		\vskip 1.5em%
		{\normalsize
			\lineskip .5em%
			\begin{tabular}[t]{c}%
				\@author
			\end{tabular}\par}%
		\vskip 1em%
		{\normalsize \@date}%
	\end{center}%
	\par
	\vskip 1.5em}
\makeatother

\usepackage[a4paper, total={17.5cm,24cm}]{geometry}

\usepackage[font=normal,labelfont=bf]{caption}

\usepackage{sectsty}

\usepackage{titlesec}
\titlespacing\section{0pt}{12pt plus 3pt minus 3pt}{1pt plus 1pt minus 1pt}
\titlespacing\subsection{0pt}{10pt plus 3pt minus 3pt}{1pt plus 1pt minus 1pt}
\titlespacing\subsubsection{0pt}{8pt plus 3pt minus 3pt}{1pt plus 1pt minus 1pt}

\titleformat{\section}{\normalfont\large\bfseries}{\thesection}{1em}{}

\titleformat{\subsection}{\normalfont\normalsize\bfseries}{\thesubsection}{1em}{}

\titleformat{\subsubsection}{\normalfont\normalsize}{\thesubsubsection}{1em}{}

\titleformat{\paragraph}[runin]{\normalfont\normalsize\itshape}{\theparagraph}{1em}{}

\usepackage{mathtools,upgreek,amsmath,amssymb}

\usepackage[utf8]{inputenc}	
\usepackage[T1]{fontenc}	
\usepackage{xcolor}		
\usepackage[linktoc=page,
colorlinks=true,	
linkcolor=blue,
citecolor=blue,
urlcolor=blue]{hyperref}     
\usepackage{booktabs} 		
\usepackage{nicefrac}		
\usepackage{microtype}		
\usepackage{lineno}		
\usepackage{float}			
\usepackage{svg}

\usepackage{siunitx}[=v2]
\usepackage{nicefrac,multirow}
\usepackage{import}

\usepackage{subcaption}
\usepackage{algorithm}
\usepackage{algpseudocode}
\usepackage[inline]{enumitem} 
\setlist[enumerate]{label*=\arabic*.}

\newcommand{\ie}{i.e.}

\DeclareMathOperator{\tr}{tr}

\newcommand{\inte}[3]{\int \limits_{ #1} #2 \; \mathrm{d} #3}

\newcommand{\nabr}{\nabla_{\ve{X}}}

\newcommand{\diffp}[2]{\frac{\partial #1}{\partial #2}}
\newcommand{\ddiffp}[2]{\dfrac{\partial #1}{\partial #2}}

\newcommand{\ve}[1]{\boldsymbol{#1}} 
\newcommand{\te}[1]{\mathbf #1}

\newcommand{\rset}{\mathbb{R}}

\newcommand{\nv}{\mathrm}
\newcommand{\comma}{\hspace{3mm} \text{,}}
\newcommand{\commaf}{\hspace{3mm} \text{,}}

\newcommand{\glmand}{\hspace{3mm} \text{and} \hspace{3mm}}
\newcommand{\glmwith}{\hspace{3mm} \text{with} \hspace{3mm}}

\newcommand{\point}{\hspace{3mm} \text{.}}

\newcommand{\pren}[1]{\prescript{n}{}{#1}}
\newcommand{\prenc}[1]{\prescript{n\!}{}{#1}}

\newcommand{\psiv}{\psi^\mathrm{vol}}
\newcommand{\psii}{\overline \psi}
\newcommand{\psiinn}{\psii^\nv{PANN}}
\newcommand{\phinn}{\phi^\nv{PANN}}

\newcommand{\piso}{\overline P}

\newcommand{\fiso}{\overline{\te F}}
\newcommand{\ciso}{\overline{\te C}}

\newcommand{\omref}{\varOmega_{0}}
\newcommand{\omegadot}{\dot{\omega}}
\newcommand{\domref}{\partial \varOmega_{0}}
\newcommand{\domrefu}{\partial \varOmega_{0,\hat{\ve u}}}
\newcommand{\domreft}{\partial \varOmega_{0,\hat{\ve T}}}

\newcommand{\veZero}{\textbf{\textit{0}}}

\newcommand{\rs}{\mathbb{R}}
\newcommand{\rsnn}{\rs_{\geq 0}}
\newcommand{\rsp}{\rs_{> 0}}
\newcommand{\so}{\mathbb{SO}(3)}

\newcommand{\tes}{\mathcal{T}^2}
\newcommand{\tesp}{\mathcal{T}^2_+}
\newcommand{\tesu}{\mathcal{T}_\nv{u}^2}
\newcommand{\tess}{\mathcal{T}_\nv{s}^2}

\newcommand{\pp}{\tilde p}

\usepackage{multicol}

\DeclareMathOperator{\cof}{cof}

\newcommand{\ici}{\overline{I}_1}
\newcommand{\iici}{\overline{I}_2}

\newcommand{\omntr}{\prescript{n,\nv{trial}}{}{\omega}}

\definecolor{pltred}{RGB}{214,39,40}
\definecolor{darkgray176}{RGB}{176,176,176}
\definecolor{pltorange}{RGB}{255,127,14}
\definecolor{darkslategray51}{RGB}{51,51,51}
\definecolor{pltgreen}{RGB}{44,160,44}
\definecolor{lightgray204}{RGB}{204,204,204}
\definecolor{pltblue}{RGB}{31,119,180}

\usepackage{layouts}
\DeclareEmphSequence{\bfseries}

\usepackage{mdframed}

\usepackage{amsthm}

\theoremstyle{definition}

\theoremstyle{remark}
\newtheorem{remark}{Remark}

\title{Precise, efficient and flexible modeling of crystallizing elastomers based on physics-augmented neural networks}
\begin{document}

	\author[$\diamond$,a]{Konrad Friedrichs}
	\author[$\diamond$,a]{Franz Dammaß}
	\author[a]{Karl A. Kalina}
	\author[,a,b]{Markus Kästner\thanks{Contact: \texttt{markus.kaestner@tu-dresden.de} }}
	\affil[$\diamond$]{These authors contributed equally.}
	\affil[a]{Institute of Solid Mechanics, TU Dresden, Germany}
	\affil[b]{DCMS -- Dresden Center for Computational Materials Science, Dresden, Germany}
	\date{}
	\maketitle

	\begin{abstract}
	We propose a precise and efficient physics-augmented neural network (PANN) to model strain-induced crystallization in rubbery polymers. {  We demonstrate that the model can be flexibly employed for both unfilled and filled natural rubber~(NR).}
	The approach is based on a two potential framework, similar to the concept of generalized standard materials (GSMs). To describe the material behavior, neural network-based free energy and dissipation potentials are employed.
	The evolution of crystallinity is derived from the two potentials.
    { To ensure boundedness of the crystallinity, a novel constrained GSM-type evolution problem is proposed. To this end, two additional Lagrange multipliers together with the corresponding Karush-Kuhn-Tucker conditions are introduced.    
    As a result, it is guaranteed that crystallinity can be interpreted as a variable of concentration type.}
    The neural network-based potentials ensure all physically desirable properties by construction. Most importantly, objectivity, material symmetry and thermodynamic consistency are automatically fulfilled.  
	In addition, an alternative derivation of the governing model equations in time-discrete form is presented based on an incremental variational framework, which also serves as the basis for a finite element implementation.
	We demonstrate the predictive capability of the PANN using three different experimental data sets from literature, considering both stress and crystallinity evolution at material point level as well as the corresponding field distributions in a notched specimen.
    Moreover, we show that model parameterization is also possible when experimental crystallinity data is not available, still enabling suitable stress predictions.
		
	\vspace{5mm}
	\noindent
	\textbf{Keywords: }Strain-induced crystallization~\textendash~Incompressibility~\textendash~Generalized standard materials~\textendash~Physics-augmented neural networks~\textendash~Constitutive model~\textendash~Finite deformations
	
	\end{abstract}
	
	\section{Introduction}
	\label{sec:intro}
	
	Elastomers find application in various technical fields, spanning from uses in tires, dampers, and membranes to medical equipment~\cite{heinrich2024}. These materials are characterized by their ability to undergo large deformations and can generally be assumed incompressible to a good approximation~\cite{boyce2000,ward2013}.
	On small scales, the deformation of rubbery polymers is primarily governed by entropic elasticity, that is, in simplified terms, by the reversible decoiling of polymer chains.
	Nevertheless, enthalpic elasticity, i.e., roughly speaking, stretching of intermolecular and covalent bonds~\cite[Sect.~8.3]{roesler2019}, can also be relevant, see also~\cite[p.~29]{dedova2020}.
	In addition, elastomers can exhibit inelastic and dissipative phenomena.
	
	Due to their outstanding mechanical properties, the use of natural rubber (NR)-based materials is particularly appealing, as they combine large deformability with high stiffness and fracture resistance~\cite{mars2004,bruening2014,xiang2022}.
	These properties are largely attributed to the self-reinforcing phenomenon of strain-induced crystallization (SIC), which occurs in regions of high stretch, for instance for 
     {axial stretches $\lambda \gtrsim 3$ in the case of uniaxial tensile loadings}~\cite{huneau2011,candau2014,rault2006a}.
	Upon unloading, the crystalline regions dissolve, reaching the fully amorphous state  {under uniaxial tensile loading for axial stretches $\lambda \lesssim 2$}~\cite{rault2006a}.
	Experimental results suggest that the phenomenon originates from the deformation-induced alignment of polymer chains, leading to the formation of crystallites with a volume in the order of $\SI{100}{nm^3}$~\cite{chenal2007} and characteristic lengths of about~$\SI{10}{nm}$~\cite{candau2014a}.
	SIC is a phenomenon of inelasticity, i.e., in experiments, it manifests as hysteresis in the stress-strain relation.
	In comparison, particularly for unfilled NR, viscous effects are significantly less pronounced, and the material behavior exhibits only minor rate-dependence~\cite{marchal2006}.
	
	Although the phenomenon has been discovered more than 100 years ago~\cite{katz1925}, many aspects of SIC are still subject of current research. For instance, the influence of multiaxiality of deformation on SIC is not completely clear, cf.~\cite{bruening2014,chen2019,khiem2022a}.
	One important reason for this is the fact that the experimental investigation of SIC requires highly sophisticated setups.
    { 
    Most commonly, the effect is analyzed based on X-ray diffraction experiments, see~\cite{trabelsi2002,toki2003,rault2006,rault2006a,chenal2007,huneau2011,candau2014a,bruening2014,xiang2022}, among others.
	The key idea is to exploit the fact that SIC leads to local variations in electron density, which can be detected in terms of additional X-ray scattering centers. The scattering intensity can then be linked to the degree of crystallinity.
	Both the formation and the melting of crystallites are accompanied by changes in temperature.
	Based on this phenomenon, indirect strategies to characterize the evolution of crystallinity have been proposed in recent years~\cite{lecam2018,khiem2022a}, circumventing the need for X-ray scattering experiments.
	The key idea is to compute the heat source associated with SIC based on the temperature field measured via infrared thermography.
	While this strategy requires less complex experimental setups and eases experimental procedures, it requires additional assumptions regarding the constitutive behavior.
    }
		
	For technical applications, filled elastomers are of predominant interest, as filler particles considerably enhance stiffness and toughness, see e.g.,~\cite{kobayashi2015}.
	The presence of fillers significantly influences the mechanics of the polymer~--~both at small scales and its effective response at the continuum level.
	For instance, specifically considering SIC, filler particles move the onset of SIC to lower stretches, which can be attributed to the effect that they act as nucleation centers~\cite{rault2006}.
	Apart from that, viscous phenomena and the Mullins effect~\cite{mullins1969} are typically notably more pronounced in filled materials.
	
	\subsection{Models of strain-induced crystallization}
	
	To describe the macroscopic behavior of crystallizing elastomers, continuum mechanical material models are required.
	In recent years, various efforts have been made to construct such material models, which are most commonly based on statistical mechanics models of entropic elasticity.
	Until now, most of these approaches are concerned with unfilled rubber.
    The starting point for formulating these models usually involves statistical considerations regarding the conformations of individual polymer chains and possible restrictions resulting from the presence of other chains and crosslinks.
	These approaches are combined with a polymer network model, representing a macroscopic material point by a representative set of nanoscopic polymer chains. 
	Based on this network model, the kinematic quantities on chain level can be computed from the macroscopic deformation tensors
	and conversely, macroscopic quantities on the continuum level, most importantly the stress tensor and a crystallinity, can be computed from the quantities on chain level.
	For this purpose, many specific models of crystallizing elastomers adopt a network model of the micro-sphere type, see~\cite{miehe2004}.
	To this end, a macroscopic material point is identified with a spherical representative network of concentrically oriented polymer chains, considering a finite number of representative directions.
	While these concepts have proven to be effective, they are associated with substantial computational effort.
	
	Most of the proposed models consider quasi-static and isothermal conditions~\cite{kroon2010, mistry2014, guilie2015, rastak2018, nateghi2018, arunachala2021}. 
	However, different approaches to describe crystallinity are proposed~--~ either scalar quantities~\cite{mistry2014} or representations, which explicitly resolve the spatial orientation of crystallites~\cite{kroon2010, guilie2015, rastak2018, nateghi2018, arunachala2021} are considered.
	In addition, the models vary in their hypotheses regarding the kinetics of crystallization and in their assumptions about the affinity of chain deformation.
	Specifically, a non-scalar crystallinity variable is not compatible with the non-affinity approach of~\cite{miehe2004}, and  alternative formulations have been introduced to circumvent this limitation.
	In~\cite{kroon2010}, an ad-hoc approach for the non-affine deformation map is proposed, whereas \cite{nateghi2018, guilie2015} rely on solely affine network deformations. 
	Alternatively, in \cite{rastak2018} and \cite{arunachala2021} the maximal advance path constraint (MAPC)~\cite{tkachuk2012} is employed for scale transitioning, which is derived based on the effective macroscopic behavior of a multitude of consecutive microscopic polymer chains. This approach, however, goes along with a very high computational effort.
	Evolution of crystallinity is predominantly defined based on the derivative of a free energy potential with respect to the degree of crystallinity~(DOC) \cite{rastak2018, mistry2014, guilie2015}, whereas an Arrhenius-type equation is proposed in \cite{kroon2010} and a dissipation potential is formulated in \cite{nateghi2018}.
	Different from the aforementioned approaches, the affine model of~\cite{loos2020a,loos2021a,loos2021} explicitly distinguishes between three phases --~the crystallized phase as well as crystallizable and non-crystallizable amorphous phases.
	The model is formulated based on a phenomenological free energy potential of hybrid Gibbs and Helmholtz type, and the concept of representative directions~\cite{freund2010}.
	
	Apart from the aforementioned models of unfilled NR,
	filled systems are considered in \cite{dargazany2014,dargazany2014a}, and a
	thermo-mechanical model extensions have been presented~\cite{khiem2022}.
     {Note that, in general, statistical mechanics-based models developed for filled systems are not suited to model the behavior of unfilled NR, and vice versa.}
	Thermo-mechanical models of unfilled NR have been proposed in~\cite{khiem2018}, following analytical network-averaging, and ~\cite{behnke2018}, considering the unit-cube approach for the macro-nano transition and developing the extended tube model~\cite{kaliske1999} further.
	
	\subsection{Data-driven constitutive modeling}
    When applied to soft materials that show a highly nonlinear and inelastic behavior, classical models, as reviewed above for modeling strain-induced crystallization, are often not sufficiently accurate and may  {require modifications to the equations} 
    if applied to new experimental data. To overcome these limitations, machine learning approaches, in particular neural networks (NNs), have emerged as powerful tools for constitutive modeling \cite{dornheim2023,fuhg2024}. These data-driven methods offer flexibility to capture complex material responses and automate the process of constitutive modeling.%
    \footnote{Note that alternative data-driven modeling strategies, e.g., spline-based approaches~\cite{wiesheier2024,moreno-mateos2025,wiesheier2026} may also be considered.}

	In the pioneering work \cite{ghaboussi1991}, feedforward neural networks (FNNs) are introduced to model hysteresis under uniaxial and multiaxial stress, utilizing input data from multiple prior time steps to account for history-dependent material behavior. 
	While NN-based constitutive modeling gained some attention in the 1990s, its progress stalled until the recent advancements in machine learning and computational efficiency revitalized interest in data-driven techniques, as reviewed in \cite{Bock2019,Liu2021,dornheim2023,fuhg2024}.
	A key innovation in this field has been the integration of physical principles into NNs, known as 
	{mechanics-informed} \cite{asad2022}, 
	{physics-augmented} \cite{klein2022,linden2023}, 
	{physics-based} \cite{aldakheel2025,baktheer2025}, 
	{physics-constrained} \cite{kalina2023},
	{physics-informed} \cite{raissi2019,henkes2022,bastek2023}, 
	or {thermodynamics-based} \cite{masi2021} approaches. 
	These methods incorporate physics either explicitly, through problem-specific architectures \cite{kalina2022,linka2021}, or implicitly, via tailored loss functions \cite{Rosenkranz2023a,weber2023,geiger2025}, thus enabling the use of sparse data for training and improved extrapolation, cf.~\cite{linden2023,masi2021,masi2024}.
	For modeling {elasticity}, many studies employ neural networks using the hyperelastic potential as output and scalar-valued invariants of the deformation tensor as inputs \cite{linden2023,klein2022,kalina2022,thakolkaran2022,linka2021,bahmani2024,benady2024,peirlinck2024}. 
	For this purpose, the concept of Sobolev training, i.e., calibrating networks representing strain energies using only stress-strain data by incorporating gradients of the network  {with respect to} the input in the objective function, is particularly effective \cite{Czarnecki2017,Vlassis2020}. 
	Additionally, polyconvexity of the neural networks \cite{klein2022,tac2022a,chen2022,linden2023,bahmani2024} has been shown to enhance extrapolation capabilities \cite{linden2023,kalina2024} and ensure rank-one convexity and,  {if the neural network based potentials are sufficiently smooth (twice continuously differentiable with respect to deformation)}, ellipticity \cite{Ebbing2010,Schroder2010}. 
	To ensure polyconvexity, fully input convex neural network (FICNN) architectures developed by~\cite{amos2017} are often used.
	Recently, an architecture based on the multiplicative split of the deformation gradient has been proposed, which enables the hybrid simulation of rubber fracture combining physics-augmented neural networks (PANNs) and classical fracture phase-field~\cite{dammass2025a,dammass2025}.
    In addition, an insightful data-adaptive approach deepening the understanding of the role of multiaxial deformations in elastomer fracture is presented in \cite{moreno-mateos2025}.
	
	  For inelastic behavior, numerous NN models proposed in the literature use internal variables and embed the networks within physically sound frameworks.
	Small strain elasto-plastic models are proposed in \cite{masi2021,vlassis2021,malik2022}, though some approaches, such as \cite{masi2021}, only weakly enforce thermodynamic consistency through loss terms and require prior knowledge of internal variables. 
	More recent approaches~\cite{meyer2023,fuhg2023} integrate thermodynamic consistency directly into the model and solve the evolution equations during training, eliminating the need for prescribed internal variables. 
	Extensions to finite strain elasto-plasticity are presented in \cite{boes2025,jadoon2025}.
	For {viscoelasticity}, the work \cite{abdolazizi2023} presents a finite strain model that builds on a generalized Prony series.
    Another approach, that was introduced in \cite{huang2022}, is to embed NNs in the generalized standard materials (GSMs) framework~\cite{biot1965,halphen1975,germain1998}, ensuring thermodynamic consistency via a dissipation potential. Similar strategies can be found in~\cite{asad2023,Rosenkranz2023a,rosenkranz2024a,holthusen2024,flaschel2025}, where some achieve training without any knowledge of the internal variable, as opposed to \cite{huang2022}, specifying only the number of internal variables.
	The most recent extensions of these approaches are concerned with the modeling of finite-strain viscoelasticity, also considering anisotropy in~\cite{holthusen2025}, and incompressible viscoelasticity~\cite{kalina2025a}.
	
	\subsection{Objectives and structure of this contribution}
	This work presents a novel NN-based modeling approach for SIC in a thermodynamically consistent way. 
	The contribution of this work is twofold. 
	First, we develop a tailored modeling framework based on GSMs, which allows to constrain the interval of admissible crystallinity values.
	Second, following the PANN approach, we employ NNs to describe the required potentials, making use of their strong expressivity.
	Accordingly, a physically sound, accurate and flexible representation of the constitutive behavior is achieved. Moreover, the novel model is also {  shown to be substantially more computationally efficient than approaches from statistical mechanics, reducing the computational effort by factors greater than 10, see Sec.~\ref{sec:calibr_unfilled}}.    
	The model proposed in this work is formulated from a macroscopic point of view, independently of the presence of filler particles, in contrast to micro-mechanical models that explicitly account for their effects. 
    We demonstrate the capability to calibrate the PANN solely relying on stress-strain data while still obtaining an internal variable from which the DOC can be approximated. Furthermore, we demonstrate that integrating the PANN into FE simulations is straightforward.
	
	The remaining paper is structured as follows. In Sec.~\ref{sec:framework}, the general model framework is developed and the governing equations for a finite element implementation are derived.
    Sec.~\ref{sec:PANN} introduces the PANN to describe the constitutive behavior.
	We examine the model, calibrated on experimental stress-strain-crystallinity data, in Sec.~\ref{sec:eval_exp_data} and provide a numerical example. 
    Further details on the extrapolation behavior of the model together with an additional validation against experimental data are given in the Appendix.
        
	\noindent\textbf{\textit{Notation.}} Within this paper, italic symbols are used for scalar quantities ($J, \psi$) and bold italic symbols for first-order tensors ($\boldsymbol{u} \in \mathcal{T}^1$). 
    For second-order tensors, bold non-italic letters ($\mathbf{P}, \boldsymbol{\sigma} \in \mathcal{T}^2$) are used. Fourth-order tensors are specified by blackboard bold symbols ($\mathbb A \in \mathcal{T}^4 $). 
	We consider a Cartesian coordinate frame in the three-dimensional Euclidean space,
	with orthonormal basis~$\{\ve e_1, \ve e_2, \ve e_3\}$ and origin~$O$ so that tensors can be identified with their coordinates, e.g., $\mathbb A = A_{ijkl} \ve e_i \otimes \ve e_j \otimes \ve e_k \otimes \ve e_l$.
	Transpose and inverse of $\mathbf{t} \in \mathcal{T}^2$ are given by $\mathbf{t}^\top$ and $\mathbf{t}^{-1}$, respectively. Additionally, $\mathrm{tr}\,\mathbf{t}, \, \det \mathbf{t}, \, \mathrm{cof}\,\mathbf{t} = \det(\mathbf{t})\, \mathbf{t}^{-\top}$ are used to indicate trace, determinant and cofactor, respectively.
	A second-order tensor $\mathbf{d}$ that is represented by a diagonal matrix with $a_1, a_2, a_3$ on its main diagonal is written as $\mathbf{d} = \mathrm{diag}(a_1, a_2, a_3)$. Furthermore, the sets of $N$-th order tensors are denoted as $\mathcal{T}^N$, $N \in \mathbb{N}_{>0}$, and 
	$
	\mathcal{T}^2_\text{s} = \left\{ \mathbf{t} \in \mathcal{T}^2 \,\middle|\, \mathbf{t} = \mathbf{t}^\top \right\}, \quad
	\mathcal{T}^2_+ = \left\{ \mathbf{t} \in \mathcal{T}^2 \,\middle|\, \det \mathbf{t} > 0 \right\}, \quad
	\mathcal{T}^2_\text{u} = \left\{ \mathbf{t} \in \mathcal{T}^2 \,\middle|\, \det \mathbf{t} = 1 \right\}.
	$
	Moreover, the special orthogonal group is given by $(\mathbb{SO}(3), \cdot)$ with 
	$
	\mathbb{SO}(3) = \left\{ \mathbf{Q} \in \mathcal{T}^2 \,\middle|\, \mathbf{Q}^{-1} = \mathbf{Q}^\top \wedge \det \mathbf{Q} = 1 \right\},
	$
	and the inner product $\cdot : (\mathbf{C}, \mathbf{D}) \mapsto \mathbf{C} \cdot \mathbf{D} = C_{kl} D_{li} \, \mathbf{e}_k \otimes \mathbf{e}_i$. Therein, $\mathbf{e}_k \in \mathbb{R}^3$ and $\otimes$ denote the $k$-th Cartesian standard basis vectors and the dyadic product, with the Einstein summation convention applied. The second order identity tensor is denoted as $\mathbf{I}$. The space of square integrable functions, which are maps between the Banach spaces $X$ and $Y$, is denoted as $\mathbb{L}^2(X; Y)$, and $\mathbb{H}^1(X; Y)$ is the first-order Sobolev space of $\mathbb{L}^2(X; Y)$ functions whose (weak) derivatives are also in $\mathbb{L}^2(X; Y)$. 
	The symbols $\nabr \ve \bullet = \diffp{\bullet_I}{X_J} \ve e_I \otimes \ve e_J$ and $\nabr \ve \cdot \bullet = \bullet_{I,I}$ are used for the gradient and divergence, respectively, of a function $\bullet$.
	Finally, the derivative of a quantity $q$  {with respect to} time is written as $\dot{q}$, and quantities related to the isochoric portion of deformation are marked by an overbar, e.g., $\overline{\psi}$.
    For reasons of readability, the arguments of functions are usually omitted within this work.

	\section{A two potential framework to model strain-induced crystallization}
    \label{sec:framework}
	In this section, we present the general continuum mechanical framework to model SIC.
	We follow the concept of generalized standard materials, and define the evolution of crystallinity based on the free energy and a dissipation potential.
	Further, we prove the thermodynamic consistency of the proposed model.
	In addition, we present a derivation of the model equations based on an incremental variational formulation, which can serve as a basis for finite element analyses.
	
	\subsection{Primary fields}
	\label{sec:hyper_fund}
	Consider a compact and connected set $\omref \subset \rs^3$ with piecewise continuous boundary~$\domref$ as the reference configuration of a solid.\footnote{Since the boundary of the domain is considered to be included in $\omref$, \ie, the set is not open, we shall assume that all field quantities can be smoothly extended over the boundary $\domref$ to an open set $\varXi \supset \omref$. This renders the computation of derivatives on $\domref$ possible, among other concepts which are technically only valid for open sets.}
	Furthermore, let $T \subset \rset$ the time interval of interest, and $\ve X \in \omref$ the coordinate of a material point in the reference configuration.
	The motion is then defined as 
	\begin{equation}
		\ve \chi : \omref \times T \longrightarrow \rset^3 \commaf \hspace{3mm} (\ve X, t) \longmapsto \ve \chi(\ve X,t) = \ve x(\ve X, t)
		\label{eq:chi}
		\commaf
	\end{equation} 
	for which sufficient smoothness is assumed.
	The displacement of a material point is defined as	$\ve u = \ve x - \ve X$,
	and the deformation gradient $\te F \in \tesp$ and its determinant $J \in \rsp$ are given by
	\begin{equation}
		\te F = 
		\nabr \ve \chi
		\qquad \text{and} \qquad J= \det \, \te F \commaf
	\end{equation}
	with $\nabr \ve \chi$ denoting the gradient of $\ve \chi$ with respect to the reference coordinate $\ve X$.
	Following Flory~\cite{flory1961}, with $\te I \in \tes$ denoting the identity, $\te F$ can be multiplicatively decomposed into 
	\begin{equation}
		\te F = \te F^\nv{vol} \cdot \fiso
		\quad \glmwith \quad \te F^\nv{vol} = J^{1/3} \, \te I
		\quad \glmand \quad \fiso = J^{-1/3}\, \te F
		\commaf
		\label{eq:flory}
	\end{equation}
	such that $\fiso\in \tesu$, i.e., $\det \fiso =1$.
	In order to quantify the deformation of the material, we introduce the symmetric positive definite right Cauchy-Green deformation tensor $\te C = \te F^\top \cdot \te F \in \tess \cap \tesp$.
	Similarly, we define its isochoric analogue by
	\begin{equation}
		\ciso  = \fiso^\top \cdot \fiso = J^{-2/3} \te C\comma
		\label{eq:cbar}
	\end{equation}
    such that $\ciso  \in \tess \cap \tesu $.
	Formulating constitutive relations by means of invariants, rather than in the components of the tensorial quantities allows to ensure objectivity and material symmetry in a straightforward manner.
	Therefore, we introduce the first and second principal invariants of $\ciso$, 
	\begin{equation}
		\ici = \tr \ciso 
		\quad \glmand \quad 
		\iici = \tr \cof \ciso \point 
		\label{eq:prinInv}
	\end{equation}
	For details on admissible invariant values and the role of the invariants in NN-based constitutive modeling, we refer to our previous work \cite{dammass2025}.\par
	The rubber materials investigated throughout this paper are assumed to be incompressible, to a good approximation. To describe their state, we introduce the field variable
	\begin{equation}
		\pp : \omref \times T\longrightarrow \rs
		\quad,\quad
		\left(\ve X, t\right) \longmapsto \pp\left(\ve X, t\right) \comma
		\label{eq:pfield} 
	\end{equation} 
	of pressure type. Moreover, we define the degree of crystallinity~(DOC)
	\begin{equation}
		\label{eq:omegafield} 
		\omega : \omref \times T \longrightarrow \left[0, 1\right]
		\quad,\quad
		\left( \ve X, t \right) \longmapsto \omega\left(\ve X, t\right) \comma
	\end{equation}
	which represents the fraction of crystalline mass.
	We assume that all primary fields are continuously differentiable with respect to time.
	
	\subsection{Free energy and stress} 
	\label{sec:energy_stress}
	
	Together with the deformation gradient $\te F$, $\pp$ and $\omega$ can be regarded as independent state variables describing the physical state of a material point.
	Accordingly, we define the Helmholtz free energy by a smooth mapping
	\begin{equation}
		\psi : \tesp \, \times \, \rs \, \times \, \left[0, 1\right]
		\longrightarrow \rs \quad,\quad 
		(\te F, \pp, \omega) \longmapsto \psi(\te F, \pp, \omega) \point
		\label{def:psi-gen}
	\end{equation}
	We employ the volumetric-isochoric split of deformation according to~\eqref{eq:flory} and specify
	\begin{equation}
		\psi(\te F, \pp, \omega) = \psii (\fiso, \omega) + \psiv(J, \pp) \quad \mathrm{with} \quad \psiv(J, \pp) = \pp \, (J-1)
		\label{def:psi-split}
		\commaf
	\end{equation}
    accounting for incompressibility.
	Furthermore, we define the isochoric portion of free energy by
	\begin{equation}
		\psii : \tesu \times \left[0, 1\right] \longrightarrow \rs
		\quad,\quad 
		(\fiso, \omega ) \longmapsto \psii (\fiso, \omega)
		\label{def:psi-iso}
		\comma
	\end{equation}
	to describe changes of free energy resulting from isochoric deformations and crystallization, cf.~\cite{dammass2025a, dammass2025}.
	For the specification of the potential $\psii$, we impose two basic requirements. 
	First, we demand material objectivity, i.e., $\psii(\te Q\cdot\fiso,\omega) = \psii(\fiso,\omega)\,\forall\,\te Q\in\so$ and isotropy, i.e., $\psii(\fiso\cdot\te Q,\omega) = \psii(\fiso,\omega)\,\forall\,\te Q\in\so$, cf.~\cite[Sect.~7]{haupt2002}.%
	\footnote{Extensions to anisotropic material behavior may be established, following, for instance, the approach of~\cite{kalina2025}.}
	Second, we assume the free energy to be zero in the absence of isochoric deformation and crystallinity, i.e., $\psii\left( \fiso = \te I , \omega = 0 \right) = 0$.

	Based on the aforementioned definition of the free energy, we introduce the first Piola-Kirchhoff stress $ \te P \in \tes$ and the crystallinity driving force $\tau\in\mathbb R$ by
	\begin{align}
		\te P &= \diffp{\psi}{\te F} 
		= \diffp{\psii(\fiso, \omega)}{\te F} + \pp \, J \, \te F^{-\top}
		\label{eq:PK1} \quad \text{and}\\
        \tau  &= - \diffp{\psi}{\omega} \quad .
        \label{eq:tau}
	\end{align}
	Note that, due to the volumetric-isochoric split considered in this work and considering exact incompressibility, the hydrostatic pressure $\mbox{$p = - {1}/({3 \, J}) \, \te P : \te F^\top$} $ can be identified as $p = - \pp$, cf.~\cite[App.~A.1]{dammass2025a}.
     {In contrast, the stress contribution related to isochoric deformations
    \begin{equation}
    \te \piso = \dfrac{\partial \psii}{\partial \te F}
    \end{equation}
    is free of contributions related to pressure.}
    In line with~\cite{rosenkranz2024a, miehe2011a}, we also refer to $\tau$
	as the force thermodynamically conjugate to the crystallinity.
	
	\subsection{Dissipation potential and crystallinity evolution}
	\label{sec:dissipation_pot}
	For the definition of the constitutive model, we follow the framework of generalized standard materials.
	To describe the inelastic effects in crystallizing rubber, we define the smooth dissipation potential
	\begin{equation}
		\phi : \rs \, \times \, \tesu
		\longrightarrow \rsnn \quad,\quad 
		(\omegadot, \fiso) \longmapsto \phi(\omegadot, \fiso) \comma
		\label{def:phi-gen}
	\end{equation}
	depending on the current isochoric deformation and the rate of crystallinity.%
	\footnote{Note that an explicit dependence of the dissipation potential on the degree of crystallinity $\omega$ is not considered in this work.}
	Similarly to the free energy, we presume that the dissipation potential guarantees objectivity and isotropy, i.e., $\phi( \omegadot, \fiso \cdot \te Q) = \phi(\omegadot, \te Q \cdot \fiso) = \phi(\omegadot, \fiso) \ \forall \ \te Q \in \so$.
	
	In addition, we define three essential properties of the potential $\phi$, intended to ensure that the material model fulfills the dissipation inequality by construction, see Sec.~\ref{sec:thermodynamic_consistency}.
	First, $\phi$ is required to be convex in $\omegadot$.
	Second, we demand $\phi(\omegadot = 0, \fiso) = 0$.
	Third, we presume that $\diffp{\phi}{\omegadot} (\omegadot = 0, \fiso) = 0$ holds, which can be interpreted as vanishing of the thermodynamically conjugated force $\hat{\tau} = \diffp{\phi}{\omegadot} \in \rs$ to crystallinity evolution for $\omegadot = 0$.
	
	Based on the two potentials $\psi$ and $\phi$ and following the formalism of GSMs~\cite{biot1965}, we define the evolution of crystallinity by the ordinary differential equation
	\begin{equation}
		\diffp{\psi}{\omega} + \diffp{\phi}{\omegadot} = \mu_0 - \mu_1
		\label{eq:biot}
	\end{equation}
	together with the Karush-Kuhn-Tucker~(KKT) conditions 
	\begin{align}
		- \omega \leq 0 \quad \land \quad &   \mu_0 \geq 0 \quad \land \quad   \omega   \mu_0 =0 \label{KKT0} \comma \\ 
		\omega -1 \leq 0 \quad \land \quad &  \mu_1 \geq 0 \quad \land \quad (  \omega -1)   \mu_1 =0 
		\commaf
		\label{KKT1}
	\end{align}
	in which $\mu_0, \mu_1 \in \rsnn$ can be understood as Lagrange multipliers that serve for ensuring $\omega \in [0, 1]$, i.e.,  crystallinity must remain between 0 and 1.%
    \footnote{ {This extension of the GSM formalism by means of $\mu_0$, $\mu_1$ and the corresponding KKT conditions is motivated from inequality-constrained optimization, where Lagrange multipliers and KKT conditions arise naturally. Whereas the time-continuous form of the evolution equation is a postulate, a time-discrete counterpart can be derived from an incremental variational principle, see Sec.~\ref{sec:variation}.}}

	\begin{remark}
		The original theory of~\cite{biot1965} is concerned with an evolution equation of the form $\diffp{\psi}{\omega} + \diffp{\phi}{\omegadot} = 0$, which does not comprise additional Lagrange multipliers.
		The proposed model may therefore be referred to as a constrained GSM formalism, accounting for the boundedness of $\omega$.
		As discussed below in Sec.~\ref{sec:thermodynamic_consistency}, this extension of the GSM concept does not hinder fulfillment of thermodynamic consistency by model construction.
	\end{remark}

	\subsection{Thermodynamic consistency}
	\label{sec:thermodynamic_consistency}
	To demonstrate the compliance of the proposed framework with the second law of thermodynamics, we follow the approach of Coleman and Noll~\cite{coleman1963} and Coleman and Gurtin~\cite{coleman1967}.
    Accordingly, we demand that the Clausius-Duhem inequality
	\begin{equation}
		{\mathcal{D}}={\te P} \, : {\dot{\te F}}^\top - {\dot \psi} \geq 0 \ \quad 
		\label{eq:cdi}
	\end{equation}
	is fulfilled for all admissible values of the state variables and all admissible rates. 
	Inserting the above definitions, in particular the free energy function $\psi$ defined in \eqref{def:psi-gen}, and the evolution equations~\eqref{eq:biot}--\eqref{KKT1}, we obtain
	\begin{equation}
		\label{eq:CDU_long}
		{\mathcal{D}} = \left( {\te P} - \frac{\partial \psi}{\partial {\te F}} \right) : {\dot{\te F}}^\top 
		-  \diffp{\psi}{\pp} \, \dot{\pp}
		+ \left(\mu_1 - \mu_0\right) {\dot{\omega}}
		+ \frac{\partial \phi}{\partial {\dot{ \omega}}} {\dot{\omega}}  \geq 0 \commaf
	\end{equation}
    which must hold true for all $\dot{\te F} \in \tes, \, \te F \in \tesp, \, \dot \pp \in \rset, \, \pp \in \rset, \, \omega \in [0,1]$ and for all $\omegadot$ fulfilling \eqref{eq:biot}--\eqref{KKT1}.
	Due to the definition of the stress~\eqref{eq:PK1}, the first term vanishes. 
	The second term expands to
	\begin{equation}
		- \diffp{\psi}{\pp} \, \dot{\pp} =  - (J-1) \, \dot{\pp}=0 \quad ,
        \label{eq:p_cdi}
	\end{equation}
    which states the unimodularity of the deformation gradient due to the exact incompressibility assumption, i.e., $J=1$, and therefore $- \diffp{\psi}{\pp} \, \dot{\pp}$ vanishes.
	The third term also vanishes, i.e.,
	\begin{equation}
		\left(\mu_1 - \mu_0\right) {\dot{\omega}}=0
		\commaf
		\label{eq:cdi_mid}
	\end{equation}
	which can be proven as follows.
	We distinguish between three different scenarios, depending on the value of $\omega$.
	First, for $\omega \in (0,1)$,  \eqref{KKT0}\textsubscript{3} and \eqref{KKT1}\textsubscript{3} imply $\mu_0=\mu_1=0$, and therefore~\eqref{eq:cdi_mid} holds.
	Second, we consider $\omega = 0$. In this case, \eqref{KKT1}\textsubscript{3} implies $\mu_1=0$ and differentiating~\eqref{KKT0}\textsubscript{3} with respect to time, we obtain
	\begin{align}
		\left[  \omega  \mu_0 \right]\dot{} = \omegadot  \mu_0 +  \omega {\dot{\mu}_0} = \omegadot  \mu_0 =0 
		\point
	\end{align}
	Third, for $\omega=1$,  \eqref{KKT0}\textsubscript{3} implies $\mu_0=0$ and \eqref{KKT1}\textsubscript{3} leads to
	\begin{align}
		\left[\left(  \omega -1 \right)  \mu_1\right]\dot{} = \omegadot  \mu_1 + \left(  \omega -1 \right) {\dot{\mu}_1} = \omegadot  \mu_1 = 0 
		\point
	\end{align}
	Therefore, \eqref{eq:cdi_mid} holds for all admissible $\omega \in [0, 1]$, and as a result, \eqref{eq:cdi} reduces to the dissipation inequality
	\begin{equation}
		\mathcal{D} = \diffp{\phi}{\omegadot} \omegadot \geq 0 
		\commaf
	\end{equation}
	which has to hold true for all $\fiso \in \tesu$ and for all $\omegadot$ fulfilling \eqref{eq:biot}--\eqref{KKT1}. 
    Due to the aforementioned properties of the dissipation potential $\phi$, which is convex w.r.t $\omegadot$ and takes the minimum $\phi(\omegadot = 0, \fiso) =0$, see e.g., \cite[p. 569]{ottosen2005}, this condition is satisfied.
	
	\subsection{Incremental variational principle}
	\label{sec:variation}
	As a basis for a finite element implementation, an alternative derivation of the governing equations is presented based on a time-discrete, incremental variational principle.
    Thereby, in addition to the evolution problem of crystallinity, we also obtain the weak forms of balance of linear momentum and pressure-motion coupling. 
	To this end, we follow ideas of~\cite[Sec.~2.2]{rambausek2022} and \cite{miehe2002c, miehe2011}.
	We consider an increment $\prescript{n}{}{t} \in \{\prescript{1}{}{t}, ..., \prescript{r}{}{t}\} \approx T$, $n, r \in \mathbb N$, in the time-discrete setting, and define two incremental potentials, from which we derive the governing equations of the model.
	
	First, we define a global potential of stored energy type ${}^n \!{\varPi}$. 
	We set $\prescript{n}{}{\ve \chi}= \ve \chi(\cdot, \prescript{n}{}{t})$ and $\prescript{n}{}{\pp} = \pp(\cdot, \prescript{n}{}{t})$ 
	and assume that a (sufficiently smooth) displacement load  $\hat{\ve{u}}\left( \ve X, t \right) : \domrefu \times T \rightarrow \mathbb{R}^3$ is prescribed on the Dirichlet part of the boundary $\domrefu \subset \domref$.
	Accordingly, we define the set of motions that are admissible at time $\pren t$ by
	\begin{equation}
		{}^n\!{M} = \left\{ \prescript{n}{}{\ve \chi}
		\in \mathbb H^1(\omref; \rs^3) \big\arrowvert 
		\prescript{n}{}{\ve \chi(\ve X)} - \ve X = \hat{\ve{u}} \left( \ve X, \pren t \right) \, \forall \ve X \in \domrefu 
		\right\} 
		\point
	\end{equation}
	Likewise, we define the set of admissible variations of motion by
	\begin{equation}
		V =	\left\{ \delta {\ve \chi} \in \left.\mathbb H^1(\omref; \rs^3) \right\arrowvert \delta \ve \chi(\ve X)  = \veZero \, \forall \ve X \in \domrefu  \right\}  \point \\
	\end{equation}
	Furthermore, we assume that the Piola traction vector $\pren{\hat{\ve{T}}}$ is prescribed on $\domreft \subset \domref$ with $\domreft \cap \domrefu = \varnothing$ at time~$\prescript{n}{}{t}$, and define the potential of external loads by
	\begin{equation}
		{}^n\!\varPi^\nv{ext} = - \inte{\domreft}{\pren{\hat{\ve{T}}} \cdot \pren{\ve \chi} }{A}
		\point
	\end{equation}
	Based on these definitions, we introduce the global incremental stored energy potential as
	\begin{align}
		&{}^n\!{\varPi}: {}^n\!{M} \times \mathbb L^2(\omref; \rset) \longrightarrow \mathbb R \quad , \quad
		(\pren{\ve \chi}, \pren\pp) \longmapsto {}^n\!{\varPi}(\pren{\ve \chi}, \pren\pp) = \inte{\omref}
		{\psi\left( \pren{\te F},  \pren{\pp},  \pren{\omega}\right)}
		{V} + {}^n\!\varPi^\nv{ext} \comma
		\label{eq:pseudo-pot_stored}
	\end{align}
	for which we consider $\prescript{n}{}{\omega} = \omega(\cdot, \prescript{n}{}{t})$ fixed.%
    \footnote{  Note that, unlike \cite{miehe2002c, miehe2011}, the incremental potential is not derived by integrating a time-continuous formulation, but is instead postulated as a time-discrete analogue of the time-continuous formulation. Nevertheless, the defined incremental potential and the one obtained by following the approach of \cite{miehe2002c, miehe2011} lead to the same stationarity conditions, and therefore to the same evolution of crystallinity.}
	
	Second, we consider a local pseudo-potential $\pren \pi$, governing the evolution of the crystallinity at the material point level. 
	We assume $\prescript{n}{}{\ve \chi}$ and $\prescript{n}{}{\pp}$ fixed and approximate the rate of crystallinity by 
	\begin{equation}
		\pren{\omegadot} \approx \frac{ \pren{\omega} - \prescript{n-1}{}{\omega}}{ {}^n\!{\Delta t}}
		\commaf
		\label{eq:implEuler}
	\end{equation}
	with the time step given by ${}^n\! \Delta t =  \prescript{n}{}{t}-\prescript{n-1}{}{t}$. 
	Based on these assumptions, we define the time-discrete local pseudo-potential using an implicit Euler scheme:
	\begin{align}
		\pren{\pi}: [0, 1] \longrightarrow \mathbb R \quad , \quad
		\pren\omega \longmapsto\pren{\pi}(\pren\omega) = 
			\psi\left( \pren{\te F},  \pren{\pp},  \pren{\omega}\right) 
			+  {}^n\!{\Delta t} \,
			\phi\left(   
			{\frac{ \pren{\omega} - \prescript{n-1}{}{\omega}}{ {}^n\!{\Delta t}}},
            \pren{\te \fiso}
			\right) 
		\point
		\label{eq:pseudo-pots}
	\end{align}
	
	We derive the governing equations demanding stationarity of the global and local potentials.
	First, stationarity of the global potential, i.e.,
	\begin{equation}
		{}^n\!{\varPi} \, \longrightarrow  \, \underset{ \pren{\ve \chi} , \pren \pp \,  \in \, {}^n\!{M} \times \mathbb L^2(\omref; \rset)}{\mathrm{stat}}
	\end{equation}
	implies vanishing Gâteaux derivatives
	\begin{align}
		\label{eq:statPi_chi}
		&\mathcal{D}_{\pren{\ve \chi}}{}^n\! \varPi \left( \pren{\ve \chi}, \pren \pp \right) \left[\delta{\ve \chi}\right] = 0 \quad 
		\forall \, \delta {\ve \chi} \in V \\
		\label{eq:statPi_p}
		&\mathcal{D}_{\pren{\pp}}{}^n\! \varPi\left( \pren{\ve \chi}, \pren \pp \right)\left[\delta{\pp}\right] = 0 \quad \forall \, \delta \pp \in  \mathbb L^2(\omref; \rset) \commaf
	\end{align}
	yielding the time-discrete weak form of mechanical equilibrium,
	\begin{equation}
		\inte{\omref}{\pren{\te P}: \left( \nabla_{\ve X} \delta {\ve \chi}\right)^\top }{V} - \inte{\domreft}{\pren{\hat{\ve{T}}} \cdot \delta {\ve \chi}}{A}= 0 
		\quad \forall \, \delta {\ve \chi} \in V \commaf
		\label{eq:weakMomBal}
	\end{equation}
    wherein the definition of the first Piola-Kirchhoff stress~\eqref{eq:PK1} is used, and the weak form of pressure-motion coupling
	\begin{equation}
		\inte{\omref}{\diffp{\pren\psi^{\mathrm{vol}}}{\pren\pp} \delta  \pp}{V} = \inte{\omref}{\left( {}^n\! J -1\right)\delta\pp}{V} = 0 \quad \forall \, \delta \pp \in  \mathbb L^2(\omref; \rset)
		\point	
		\label{eq:weakPressMo}
	\end{equation}
	If we assume the field quantities to be sufficiently smooth, \eqref{eq:weakMomBal} and \eqref{eq:weakPressMo} are equivalent to the balance of linear momentum in the form
	\begin{equation}
		\nabr \cdot \prescript{n}{}{\te P} = \veZero \quad  \nv{in} \,\, \omref
		\quad \glmand \quad
		\prescript{n}{}{\te P} \cdot \ve N = \prescript{n}{}{\hat{\ve{T}}} \quad \nv{on} \,\, \domreft
		\commaf
	\end{equation}
	with $\ve N$ denoting the outer surface normal on $\domreft$
	and the incompressibility condition
	\begin{equation}
		{}^n\!{J} = 1
		\commaf
		\label{eq:incompr}
	\end{equation}
	respectively.
	Similarly, demanding stationarity of the time-discrete local pseudo-potential under the constraints \mbox{$0 \leq \pren{\omega} \leq 1$}, i.e.,
	\begin{equation}
		\pren{\pi} \, \longrightarrow  \, \underset{ \pren{\omega} \in [0,1]}{\mathrm{stat}} 
		\commaf
        \label{eq:pi_local_stat}
	\end{equation}
	we obtain the conditions
	\begin{align}
		\diffp{\psi\left( \pren{\te F},  \pren{\pp},  \pren{\omega}\right) }{\pren \omega} 
		&+ {}^n\!{\Delta t} \, \diffp{\phi\left(   
			{\dfrac{ \pren{\omega} - \prescript{n-1}{}{\omega}}{ {}^n\!{\Delta t}},  \pren{\te \fiso}}
			\right) }{\pren{\omega}} = \pren \mu_0 - \pren \mu_1 
		\commaf \label{eq:biot_discr}\\
		- \pren\omega \leq 0 \quad &\land \quad   \pren\mu_0 \geq 0 \quad \land \quad   \pren\omega   \pren\mu_0 =0 \commaf \label{eq:kkt1_discr}\\
		\pren\omega -1 \leq 0 \quad &\land \quad  \pren\mu_1 \geq 0 \quad \land \quad (  \pren\omega -1)   \pren\mu_1 =0 \label{eq:kkt2_discr}
		\commaf
	\end{align}
	i.e., time-discrete versions of the evolution equation~\eqref{eq:biot} and the Karush-Kuhn-Tucker conditions~\eqref{KKT0} and \eqref{KKT1}.%
    \footnote{To show that \eqref{eq:biot_discr} is a time-discrete analogue of~\eqref{eq:biot}, we can understand the finite difference~\eqref{eq:implEuler} as a bijective mapping $\pren{\omega} \mapsto \pren{\omegadot}(\pren{\omega})$ with inverse
    $\pren{\omegadot} \mapsto \pren{\omega}(\pren{\omegadot})$.
    Applying the chain rule and inserting~\eqref{eq:implEuler}, we obtain $\diffp{\phi(\pren \omegadot, \pren \fiso)}{\pren\omega} = \diffp{\phi(\pren \omegadot, \pren \fiso)}{\pren \omegadot} \diffp{\pren \omegadot}{\pren \omega} = \frac{1}{{}^n\! \Delta t}\diffp{\phi(\pren \omegadot, \pren \fiso)}{\pren\omegadot}$.}
	Note that the same equations can be obtained by discretization of~\eqref{eq:biot}, \eqref{KKT0} and \eqref{KKT1} following the implicit Euler approach~\eqref{eq:implEuler}.

	\subsection{Numerical aspects}
	\label{sec:FE_implementation}
	
	In the following, we elaborate on the solution scheme for the evolution of crystallinity and problem-specific aspects of the finite element implementation used for the numerical examples presented in Sec.~\ref{sec:eval_exp_data}.
	
	\subsubsection{Update of crystallinity}
	\label{sec:crystEvAlg}
	
	To compute the crystallinity values, we iteratively solve the evolution problem~\eqref{eq:biot_discr}--\eqref{eq:kkt2_discr} at material point level. To this end, we employ a predictor-corrector type algorithm, which is conceptually similar to standard approaches in elastoplasticity, cf.~\cite{wriggers2008}, and whose essential ideas are presented in the following.
	
	We consider the $n$-th increment within the discretized time interval, i.e., $\pren t \in \{\prescript{1}{}{t}, ..., \prescript{r}{}{t}\}$ and aim at computing~$\pren \omega$. We assume that $\prescript{n-1}{}{\omega}$ is known and $\pren{\te F}, \, {}^n\! \Delta t$ are given.%
	\footnote{Without loss of generality, we define the initial conditions $\prescript{0}{}{\omega} = 0, \, \prescript{0}{}{\te F} = \te I$ at $\prescript{0}{}{t} = \SI{0}{\second}$.
		Also note that, when the predictor-corrector scheme is employed at the quadrature point level to compute a finite element solution, the given value of $\pren{\te F}$ does not necessarily refer to the converged value of the corresponding increment, but rather to the value from the current global Newton iteration for the nodal values of displacement.}
	In the preliminary predictor step, we compute a trial value of crystallinity $\omntr$ assuming $\pren \mu_0 = \pren \mu_1 =0$, i.e., without enforcing the constraint~$\omntr \in [0,1]$.
	To this end, we solve the reduced form of~\eqref{eq:biot_discr},
	\begin{equation}
		\label{eq:residual}
		{}^n\! R = \diffp{\psi\left( \pren{\te F},  \pren{\pp},  \omntr\right) }{\omntr} 
		+ {}^n\!{\Delta t} \, \diffp{\phi\left(    
			{\dfrac{ \omntr - \prescript{n-1}{}{\omega}}{ {}^n\!{\Delta t}},\pren{\te \fiso},}
			\right) }{\omntr} 
		= 0
		\commaf
	\end{equation}
	for $\omntr$, employing a local Newton iteration scheme.
	If the obtained solution is indeed within the admissible interval, we assume $\pren \omega =\omntr$.
	Otherwise, in the case $\omntr \notin [0, 1]$, a corrector step follows.
	The key idea is to push the crystallinity back to the closest boundary of the admissible interval. In case of $\omntr < 0$, we assume $\pren \omega = 0$ and compute $\pren \mu_0$ from~\eqref{eq:biot_discr}, accordingly. Conversely, if $\omntr > 1$, $\pren \omega = 1$ is assumed and the corresponding value $\pren \mu_1$ is identified from~\eqref{eq:biot_discr}.
    { Subsequently, it is verified that \eqref{eq:kkt1_discr}\textsubscript{2} and \eqref{eq:kkt2_discr}\textsubscript{2}, respectively, are satisfied, i.e., that $\pren \mu_0  \geq 0$ and $\pren \mu_1 \geq 0$ hold.}%
    \footnote{ 
    For the present model, we cannot provide a formal proof that the corrector step in every scenario indeed yields two non-negative~$\pren \mu_0, \pren \mu_1 $.
    However, their non-negativity has been verified numerically, considering a representative set consisting of a large number of test state variable tuples. These also include extreme deformation states that exceed the physically relevant range, such as uniaxial deformation with axial stretches~$\lambda \in [0.02, 20]$.
    A possible strategy to guarantee that the corrector step yields non-negative Lagrange multipliers by construction of the model would be to require convexity of $\psi$ with respect to $\omega$.
    However, this model variant resulted in a poorer approximation of the experimental data compared to the proposed model and was therefore not pursued further in this work.
    }

	\subsubsection{Finite element implementation}
	
	The numerical solution of the field problem is based on a {discretized version of the weak form of equilibrium}~\eqref{eq:weakPressMo} together with the incompressibility condition~\eqref{eq:incompr} and the time-discrete evolution of crystallinity~\eqref{eq:biot_discr}--\eqref{eq:kkt2_discr}.
	{Discretization in space} is carried out based on the total Lagrangian approach, following the isoparametric finite element method:
	We approximate the reference configuration
	using a mesh of isoparametric 6-node triangular elements with quadratic shape functions.
	Throughout this work, we exclusively consider incompressible materials and assume plane stress states in the $X_1, X_2$ plane. 
	Accordingly, the deformation gradient and first Piola-Kirchhoff stress take the forms
	\begin{equation}
		\left[F_{kl}\right] = \begin{bmatrix}
			F_{11}&F_{12}&0\\
			F_{21}&F_{22} & 0\\
			0 & 0 & F_{33}
		\end{bmatrix} 
        \quad
        \text{and}
        \quad
        \left[P_{kl}\right] = \begin{bmatrix}
			P_{11}&P_{12}&0\\
			P_{21}&P_{22} & 0\\
			0 & 0 & 0
		\end{bmatrix}
		\label{eq:F_planeStress}
		\quad ,
	\end{equation}
    respectively.
	Consequently, we consider 2D displacement-based finite elements to compute the in-plane field quantities~\cite{pascon2019}. We address in-plane coordinates of tensors using Greek indices, e.g., $\alpha,\beta\in\{1,2\}$.
    Thus, we determine the out-of-plane component $F_{33}([F_{\alpha\beta}])$ from the incompressibility condition~\eqref{eq:incompr}, which, inserting~\eqref{eq:F_planeStress}, can be written as a function of the in-plane components:
	\begin{equation}
		F_{33}([F_{\alpha\beta}])=\left(F_{11} F_{22}-F_{21} F_{12}\right)^{-1}
		\label{eq:F_33_planeStress}
		\point
	\end{equation}
	The pressure-type field $\pp$ is then identified from~\eqref{eq:PK1} with the stress boundary condition
    \begin{equation}
    \label{eq:pressure_planestress}
        {}^n\! P_{33}([{}^n\! F_{\alpha\beta}],{}^n \omega) = 0 \; \Longleftrightarrow \; 	
		 \pren \pp([{}^n\!F_{\alpha\beta}],{}^n \omega) = -\pren{\piso_{33}}([{}^n\! F_{\alpha\beta}],{}^n \omega) {}^n\!{F_{33}}([{}^n\! F_{\alpha\beta}]) \point
    \end{equation}
	Based on Galerkin's method and the assembly of finite element shape functions, the infinite-dimensional function spaces ${}^n\! M$ and $V$ are approximated by finite-dimensional counterparts.
	At each time increment, the weak form of equilibrium is then solved iteratively for the nodal values of displacement. To this end, a standard nonlinear finite element solver approach based on Newton's method is employed, cf.~\cite{wriggers2008}.
	Conversely, the crystallinity is only resolved at the quadrature points of the finite elements, following the standard strategy for the numerical implementation of inelastic material models.
	To compute the crystallinity values, we iteratively solve the evolution problem~\eqref{eq:biot_discr}--\eqref{eq:kkt2_discr} locally at every integration point within each iteration of the global finite element solver.
	To this end, we employ the aforementioned predictor-corrector type algorithm.
	
	The global Newton scheme to compute the nodal degrees of freedom requires a linearization of the stress-deformation map, which must be carried out consistently with the update algorithm for crystallinity, see~\cite{wriggers2008} for more details in the context of elastoplasticity.
	
	To this end, we define the {  algorithmic} consistent material tangent~$\mathbb A\in\mathcal{T}^4$, which is a fourth-order tensor.
	For the present 2D plane stress FE model, we only consider its in-plane components $A_{\zeta \eta \nu \xi}$, cf.~\cite{pascon2019}:
	\begin{equation}
		\prenc{A}_{\zeta \eta \nu \xi} = \frac{\partial \prenc{ P}_{\zeta \eta }}{\partial \prenc{ F}_{\nu \xi}} + \diffp{\prenc{ P}_{\zeta \eta}}{\pren \omega}\diffp{\pren \omega}{\prenc{F}_{\nu \xi}} = \frac{\partial \prenc{ P}_{\zeta \eta }}{\partial \prenc{ F}_{\nu \xi}} - \diffp{\prenc{P}_{\zeta \eta}}{\pren \omega} \diffp{\prenc R}{\prenc{ F}_{\nu \xi}}\left(\diffp{\prenc R}{\pren\omega}\right)^{-1} \quad .
		\label{eq:matTan}
	\end{equation}
    Therein, the in-plane stress tensor, which accounts for the plane-stress boundary condition, follows from \eqref{eq:PK1} together with \eqref{eq:pressure_planestress} as
    \begin{equation}
        \prenc P_{\zeta \eta}([\prenc F_{\alpha\beta}],{}^n \omega) = \pren{\overline{P}}_{\zeta \eta}([\prenc F_{\alpha\beta}],{}^n \omega) - \frac{\pren{\overline{P}}_{33}([\prenc F_{\alpha\beta}],{}^n \omega)}{\prenc F_{11} \prenc F_{22} - \prenc F_{21} \prenc F_{12}} \prenc F_{\eta \zeta}^{-1} \point
    \end{equation}
	The  derivatives required in~\eqref{eq:matTan} are obtained using automatic differentiation.
	Thereby, we implicitly account for the dependence of the in-plane stress components on the out-of-plane deformation through the pressure~$\pren p$, as given by~\eqref{eq:pressure_planestress}.
	
	\section{Physics-augmented neural network-based modeling of SIC}
    \label{sec:PANN}
	To specify the isochoric free energy $\psii$ and the dissipation potential $\phi$, we follow the idea of physics-augmented neural networks~(PANNs)~\cite{klein2022,linden2023,kalina2024,dammass2025}, which allows to combine physical consistency, accuracy, computational efficiency and flexibility.
	
	\subsection{Requirements for the potentials}
	\label{sec:NN_requirements}
	
	To ensure physical consistency of the neural network-based model, we demand several properties of the potentials { defined in Sec.~\ref{sec:framework}} to be ensured by the architecture of the model, cf.~\cite{linden2023,dammass2025}.
	
	For both the isochoric free energy and the dissipation potential, we require:
	\begin{enumerate}[label=(\roman*)]
		\item Objectivity:%
		\footnote{Note that objectivity of $\psii$ implies fulfillment of the balance of angular momentum, i.e., $\te P \cdot \te F^{\top} = \te F \cdot \te P^{\top}$, see~\cite[Proposition~8.3.2]{silhavy1997}.}
		$\psii(\te Q \cdot \fiso, \omega)=\psii(\fiso,\omega)$ and $\phi(\omegadot, \te Q \cdot\fiso)=\phi(\omegadot,\fiso)\quad \forall \, \te Q \in \so  \quad \forall \, \fiso \in \tesu$ 
		\item Isotropy: $\psii(\te \fiso \cdot\te Q, \omega)=\psii(\te \fiso,\omega)$ and $\phi(\omegadot, \fiso\cdot\te Q)=\phi(\omegadot, \fiso) \quad \forall \, \te Q \in \so  \quad \forall \, \fiso \in \tesu$. 
	\end{enumerate} 
	
	In addition, to ensure compatibility with the second law of thermodynamics as discussed in Sec.~\ref{sec:thermodynamic_consistency}, the dissipation potential must provide the following properties:
	\begin{enumerate}[label=(\roman*)]
		\setcounter{enumi}{2}    
		\item Convexity  {with respect to} $\dot{\omega}$: $\dfrac{\partial^2\, \phi}{\partial \, \dot{\omega}^2 }(\dot \omega, \fiso) \geq 0 \quad  \forall \, \dot \omega \in \rset \quad \forall \, \fiso \in \tesu $   ,
		\item Zero dissipation potential in the absence of crystallinity evolution: $\phi(\omegadot = 0, \fiso) = 0 \quad \forall \, \fiso \in \tesu$ ,
		\item Non-negativity:\footnote{It is worth mentioning that the conditions (iii) $\land$ (iv) $\land$ (v) are equivalent to (iii) $\land$ (iv) and demanding a vanishing thermodynamically conjugate force to crystallinity, when no crystallinity evolution takes place, \ie, \linebreak $\hat \tau(\dot \omega=0, \fiso) = \ddiffp{\phi}{\dot \omega}( \dot \omega=0, \fiso) = 0 \quad \forall \, \fiso \in \tesu$.} 
        $\phi (\omegadot, \fiso) \geq 0 \quad  \forall \, \dot \omega \in \rset \quad \forall \, \fiso \in \tesu $.
	\end{enumerate}
	We further construct the NN representing $\psii$ such that it ensures: 
	\begin{enumerate}[label=(\roman*)]   
		\setcounter{enumi}{5}
		\item Zero energy in the absence of deformation and crystallinity: $\psii( \fiso=\te I, \omega = 0 ) = 0$
		  \item Zero stress in the absence of deformation and crystallinity: $\te \piso ( \fiso=\te I, \omega = 0 ) = \te 0$ 
		\item Zero thermodynamically conjugate force to crystallinity, in the absence of deformation and crystallinity:
		\linebreak $\tau(\fiso=\te I, \omega = 0) = -\ddiffp{\psii}{\omega}(\fiso=\te I, \omega = 0) = 0$
		\item Polyconvexity of the free energy with respect to the deformation, in the sense of \cite{ball1976}:
		There is an		
		\[
		{\mathcal F} : \rset^{3\times3} \times \rset^{3\times3} \times \rset \times [0,1] \longrightarrow \rset \cup \{\infty\}
		\, , \quad
		(\te F, \cof \te F, J, \omega) \longmapsto {\mathcal F}(\te F, \cof \te F, J, \omega) 
		\]
		such that ${\mathcal F}(\te F, \cof \te F, J, \omega) \equiv \psii(\fiso, \omega)$ and ${\mathcal F}$ is convex  {with respect to} $(\te F, \cof \te F, J)$.%
        \footnote{Note, however, that this does not imply polyconvexity of the entire SIC model. 
        { Also note that we require polyconvexity of $\psii$ independently of the specific choice for $\psiv$, cf.~\eqref{def:psi-split}. As a result, polyconvexity of $\psii$ would also be ensured if nearly incompressible or even compressible behavior were assumed.}}
		
	\end{enumerate}
	\begin{remark}
		The concept of polyconvexity has its roots in hyperelasticity, more precisely in the context of proving the existence of energy minimizers under the assumption of hyperelastic material behavior.
		In the context of neural network-based material modeling, polyconvex potentials have shown superior to unconstrained approaches regarding the capability to make plausible predictions to unseen deformation states, the robustness against noise in experimental data as well as overfitting when trained on sparse experimental data, see~\cite{linden2023,dammass2025}.
		However, it seems worth mentioning that a non-polyconvex potential does neither {a priori} violate physical laws nor {a priori} implies the prediction of unrealistic material behavior. 
    Conversely, for instance in the context of multiscale modeling and computational homogenization, enforcing polyconvexity may hinder model accuracy, see e.g., \cite{kalina2025}.
	\end{remark}

	\subsection{Definition of the potentials and network architecture}
\begin{figure}[h]
		\centering
		\includegraphics[]{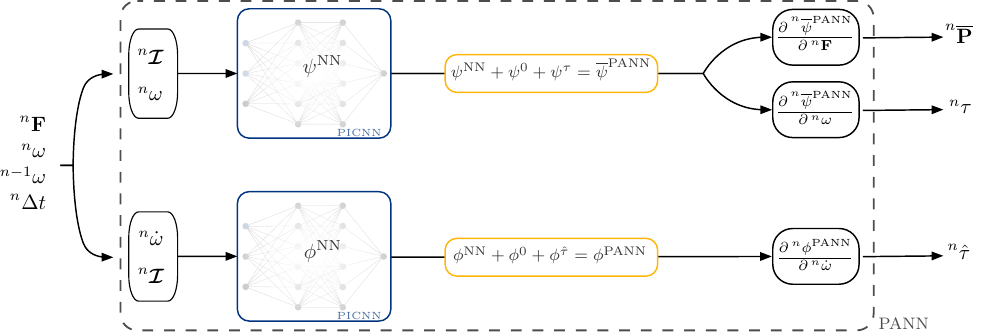}
	\caption{Schematic of the proposed PANN model for SIC. The DOC $\pren \omega$ of the current increment follows from the compliance of $\pren \tau$ and $\pren{\hat{\tau}}$ with the evolution equations \eqref{eq:biot}--\eqref{KKT1}.}
	\label{fig:PANN-modelstructure}
\end{figure}

	For the specification of the potentials $\psiinn$ and $\phinn$, we follow the approach of an invariant-based formulation \cite{kalina2022,kalina2023,dammass2025}, implying isotropy and objectivity~(i, ii).
	To this end, we consider the pair of isochoric deformation invariants
	\begin{equation}
		\boldsymbol{\mathcal{I}} = \left( \ici, \iici^{3/2} \right) \quad ,
	\end{equation}
	which are polyconvex,%
	\footnote{Note that the invariant $\iici$, different from 
		$(\iici^{3/2})$, is not generally polyconvex \cite[Lemma~2.4]{hartmann2003b}.}
	see~\cite[Lemma~2.2]{hartmann2003b}.
	Due to the choice of these invariants, zero stress in the absence of deformation is automatically fulfilled~(vii), see~\cite[Theorem~1]{dammass2025a}.
	The core ingredients to the model are two partially input convex neural networks networks~(PICNNs) with monotonic activation functions as introduced in \cite{amos2017}.
	For the dissipation potential, we consider the PICNN $\phi^{\mathrm{NN}}\left(  \dot{\omega}, \boldsymbol{\mathcal{I}}(\fiso) \right)$, which is convex with respect to $\dot \omega$.
	For the free energy, we use the PICNN $\psi^{\mathrm{NN}}\left( \boldsymbol{\mathcal{I}}(\fiso), \omega \right)$, which is convex and monotonic with respect to the invariants $\ici, \iici^{3/2}$ and thus polyconvex.
	For details on the PICNN architecture, we refer to~\cite[App.~A.1.2]{rosenkranz2024a}.
	As a result, the requirements of convexity (iii, ix) 
    are satisfied.
	To fulfill the conditions (iv, v), which require the dissipation potential and its derivative to vanish for $\dot \omega=0$, we follow the ideas of~\cite{Rosenkranz2023a,rosenkranz2024a} and define
	\begin{equation}
		\label{eq:phi_NN_corrected}
		\phinn( \dot \omega, \fiso) = \phi^{\mathrm{NN}}\left(\dot{\omega}, \boldsymbol{\mathcal{I}}(\fiso) \right) + \phi^\mathrm{0}\left( \boldsymbol{\mathcal{I}}(\fiso) \right) + \phi^\mathrm{\hat{\tau}}\left(  \dot{\omega}, \boldsymbol{\mathcal{I}}(\fiso) \right)
		\commaf
	\end{equation}
	including the two correction terms
	\begin{equation}
		\phi^\mathrm{0}( \dot \omega, \fiso) = -\phi^{\mathrm{NN}}\left(\dot{\omega}=0, \boldsymbol{\mathcal{I}}(\fiso)\right)
		\quad \text{and} \quad 
		\phi^\mathrm{\hat{\tau}} = -\frac{\partial \phi^{\mathrm{NN}}}{\partial \dot{\omega}}(\dot{\omega}=0, \boldsymbol{\mathcal{I}}(\fiso)) \, \dot{\omega} 
		\point
	\end{equation}
	For the free energy, we similarly define
	\begin{equation}
		\label{eq:psi_NN_corrected}
		\psiinn = \psi^{\mathrm{NN}}\left( \boldsymbol{\mathcal{I}}(\fiso), \omega \right) + \psi^\mathrm{0} + \psi^\mathrm{\tau} (\omega)\comma
	\end{equation}
	including the correction terms 
	\begin{equation}
		\psi^\mathrm{0} = -\psi^{\mathrm{NN}}\left( \boldsymbol{\mathcal{I}}(\fiso = \te I), \omega = 0 \right)
		\quad \text{and} \quad 
		\psi^\mathrm{\tau} = \left. -\frac{\partial \psi^{\mathrm{NN}}}{\partial \omega}\right|_{\fiso = \te I, \omega = 0} \omega
		\commaf
	\end{equation}
	which ensure fulfillment of the conditions (vi) and (viii).
	An overview on the PANN architecture is given in  Fig.~\ref{fig:PANN-modelstructure}.
	The isochoric stress $\pren {\te \piso}$ and the thermodynamically conjugate forces ${}^n\tau$ and ${{}^n\hat{\tau}}$ governing the evolution of crystallinity are computed as the derivatives of the PANN's free energy $\psiinn$~\eqref{eq:psi_NN_corrected}~and dissipation potential~$\phinn$~\eqref{eq:phi_NN_corrected}. 
	The total stress ${}^n{\te P}$ can be calculated by adding the contribution from hydrostatic pressure, see~\eqref{eq:PK1}.
	The evolution of crystallinity can be determined as discussed above in Sec.~\ref{sec:crystEvAlg}.

\begin{remark}
	To address the significantly varying magnitudes of the network inputs and expected outputs, we employ the idea of~\cite{kalina2025} and adopt non-trainable normalization layers that perform a simple linear scaling of the input and output quantities for numerical purpose. Details are omitted here, for the reason of brevity, and the reader is referred to \cite[App.~D]{kalina2025}.
\end{remark}

\begin{remark}
		Deviating from the above definition of $\boldsymbol{\mathcal{I}}$, we also present results for a PANN, which solely considers $\ici$ as deformation measure. 
		This is due to the fact that, to date, experimental data sets comprising both stress-stretch and crystallinity-stretch data are solely available for uniaxial tensile experiments.
		However, as discussed in~\cite{dammass2025}, $\iici$ only takes the smallest admissible values in uniaxial tensile experiments.		
		Thus, any dependencies on $\iici$, supposedly learned from the dataset by the PANN lack sufficient basis. As a consequence, models based on both $\ici$ and $\iici$ typically predict overly stiff responses for load cases other than uniaxial tension, as thoroughly discussed in \cite{dammass2025}.
		Conversely, models only considering $\ici$ can make plausible predictions for multiaxial deformation states.
		\label{rem:i1only}
	\end{remark}
	
	\subsection{Model calibration}
	\label{sec:NN_training}
    The terms \textit{calibration} and \textit{training} are used synonymously in the following, emphasizing that the technique presented below is not restricted to the PANN model.  
    In general, any formulation for the potentials can be applied in a similar way and calibrated accordingly.
    \subsubsection{General approach}
\begin{figure}[tb]
        \centering
        \includegraphics{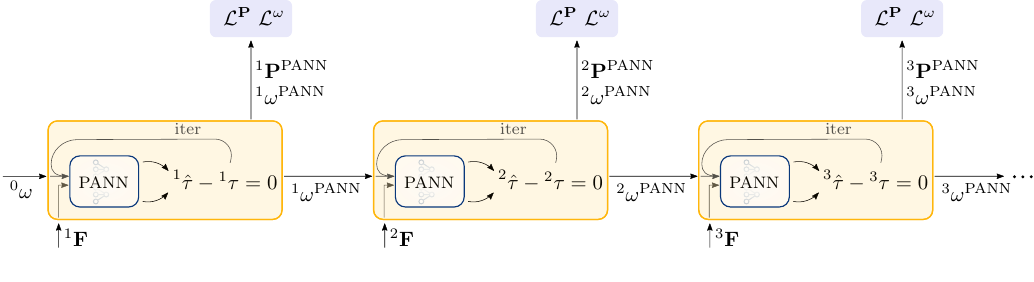}
        \caption{Training scheme of the PANN based on the constrained optimization problem defined in Eq. \eqref{eq:training_optimization}. At each time step, the current DOC is determined by iteratively solving the unconstrained evolution equation \eqref{eq:biot}, i.e., under the presumption $\mu_0 = \mu_1 = 0$. Consequently, evaluating the stress at any increment of the training sequence requires the evaluation of all preceding time steps. The illustration is based on \cite{rosenkranz2024a,kalina2025a}.}
        \label{fig:PANN_training}
    \end{figure}

	The identification of the network parameters is carried out based on experimental stress, crystallinity and deformation data from the literature.%
	\footnote{Experimental stress and crystallinity data from the literature are not necessarily synchronized in time, see e.g.,~\cite{rault2006a}. Here, the model is evaluated for deformation states with corresponding stress measurements and the crystallinity needs to be interpolated during training.}
	Training PANNs for inelastic material behavior can be approached in various manners, see~\cite{rosenkranz2024a} for an extensive discussion.
	In this work, we pursue the approach of integrating the evolution equation during training, see~\cite[Sec.~4.1]{rosenkranz2024a},
	since alternative approaches, such as using an auxiliary FNN or RNN, did not reveal to be advantageous here.
	As illustrated in Fig. \ref{fig:PANN_training}, the key idea is to solve the evolution equation~\eqref{eq:biot_discr} during the training for every moment in time for which experimental data is available, yielding the crystallinity and the stress, which are to be compared to the corresponding experimental ground-truth values. 
	For the training, we do not enforce the boundedness constraint on crystallinity, which is for two reasons. 
	First, experimental crystallinity values naturally lie within the admissible interval. 
	Second, this ensures differentiability of the objective function and thereby improves training stability.
	
	We consider standard experiments for training, where the pressure-type field $\pp$, can be identified from the boundary conditions, see~\cite{dammass2025} for details. 
	For instance, we consider uniaxial tension (UT), where the sparsity of the resulting stress tensor and associated deformation gradient,
		\begin{equation}  
			\te F^{\mathrm{UT}} = \mathrm{diag}(\lambda_1,\lambda_1^{-1/2}, \lambda_1^{-1/2})  \Longrightarrow \te P^{\mathrm{UT}}=\mathrm{diag}(P_{11}, 0, 0)
		\end{equation}
	can be exploited to identify the nonzero stress component
	\begin{equation}
		P_{11} = \piso_{11} -\piso_{33} \frac{\lambda_3}{\lambda_1} \comma
	\end{equation}
	only from $\psii$, wherein $\lambda_3 =  \lambda_1^{-1/2}$.
	
	The loss function is defined based on two ingredients, related to the stress and the crystallinity, respectively. We designate the vectors of experimentally determined values of axial stress $P_{11}$ and of crystallinity by ${\underline{\ve P}}^\mathrm{exp}$ and $\underline{\ve{\omega}}^{\mathrm{exp}}$, and the corresponding model predictions by ${\underline{\ve P}}^\mathrm{PANN}$ and $\underline{\ve{\omega}}^{\mathrm{PANN}}$.
	We consider the mean absolute error as deviation measure,\footnote{The mean squared error is commonly preferred as a performance metric in the training of neural networks. However, in the present work, the mean absolute error was found to yield more robust training behavior.}
    given by 
	\begin{equation}
		\mathrm{MAE}: (\underline{\ve a}, \underline{\ve b}) \longmapsto \frac{1}{I}\sum_{i=1}^I | a_i-  b_i| 
		\commaf
	\end{equation}
	and define
	\begin{equation}
		\mathcal{L}^{\te P} = \frac{1}{N_{\te P}} \mathrm{MAE}\left( {\underline{\ve P}}^{\mathrm{PANN}} , {\underline{\ve P}}^\mathrm{exp} \right) \quad \text{and} \quad
		\mathcal{L}^{\omega} = \frac{1}{N_\omega} \mathrm{MAE}\left( \underline{\ve{\omega}}^{\mathrm{PANN}} , \underline{\ve{\omega}}^\mathrm{exp}  \right) \point
	\end{equation}
	For numerical purpose, loss normalization is performed through $N_\bullet = \max(\bullet^{\mathrm{exp}}) - \min(\bullet^{\mathrm{exp}})$ and the network parameters are identified through constrained minimization as
	\begin{equation}
		\boldsymbol{\varTheta}^{\psi}, \boldsymbol{\varTheta}^\phi = \underset{\boldsymbol{\varTheta}^\psi \in C^\psi, \boldsymbol{\varTheta}^\phi  \in C^\phi}{\arg \min}\left( w^{\te P }\mathcal{L}^{\te P} + w^\omega \mathcal{L}^\omega \right)
        \quad \text{subject to} \quad \pren{\pi} \, \longrightarrow  \, \underset{ \pren{\omega} \in \rset}{\mathrm{stat}} 
        \label{eq:training_optimization}
	\end{equation}
	making use of the quasi-Newton optimizer SLSQP.  {Thus, the networks are calibrated through a first-order Sobolev training.}
	Therein, the admissible sets of weights and biases $ \boldsymbol{\varTheta}^\psi, \, \boldsymbol{\varTheta}^\phi$ are designated as $C^\psi, \, C^\phi$ and defined such that the convexity requirements discussed above are ensured, see \cite[App.~A.1.2]{rosenkranz2024a} for a detailed discussion of the PICNN architecture and the corresponding parameters.
    
	\subsubsection{Calibration without experimental crystallinity data}
    \label{sec:training_withoutDOC}
	
	Although crystallinity can generally be determined experimentally, this requires specialized, elaborate, and expensive experiments, see { Sec.~\ref{sec:intro}.}%
	Therefore, it can also be of interest to parameterize the models of crystallizing elastomers and predict their stress responses to unseen data without considering experimental crystallinity data.
	In the following, we perform training solely based on experimental stress data, allowing for the PANN to automatically find a suitable scalar-valued internal pseudo-crystallinity variable $q$ without the presumption that it has to strictly resemble the DOC. 
	Consistently, $q$ then takes the place of $\omega$ in all equations previously introduced.%
	\footnote{Note that, in general, internal variables serving to describe inelastic material behavior are ambiguous, cf. \cite{liu2023a, flaschel2025}.}
    The evolution problem of this pseudo-crystallinity 
    \begin{align}
        \diffp{\pren\psi}{\pren q} +& \prenc \Delta t \, \diffp{\pren\phi}{\pren{q}} = \mu_0 \\
        -\pren q \leq 0 \quad \land \quad \pren \mu_0& \geq 0 \quad \land \quad \pren q \pren \mu_0 = 0
    \end{align}
    then corresponds to the proposed constrained GSM framework, with $\pren q$ only bounded from below.
	To ease the comparison with the DOC and to ensure $\pren q \geq 0$ during training, we introduce a loss to guide the internal variable to be positive
	\begin{equation}
		\label{eq:Loss_negativity}
		\mathcal{L}^- = \frac{1}{N_\omega}\sum_n \max\left(0 \, , - \prescript{n}{}{q}^\mathrm{PANN}  \right)
		\commaf
	\end{equation}
	which takes the place of $\mathcal{L}^\omega$ in the optimization problem. Conversely, similarly to the model calibration with prescribed crystallinity, boundedness of $q$ is not enforced in a strong sense during training.
	Apart from that, training is performed similarly to what is described above.

    The implementation of the PANN model and the training workflow was realized using Python and the libraries TensorFlow and SciPy.	
	
	\section{Evaluation of the model performance through experimental data}
	\label{sec:eval_exp_data}
	In the following, we examine the performance of the proposed PANN model. For this purpose, we investigate different loading sequences, for which experimental stress, crystallinity and deformation data is available in the literature.
	First, we consider uniaxial tensile experiments for unfilled NR of~\cite{rault2006a}. 
	Second, we demonstrate how the stress responses of NR observed in~\cite{rault2006a} can be captured even if no crystallinity data is available from experiments, based on an automatically discovered pseudo-crystallinity.
	Third, to emphasize the versatility of our model, we also present the capability to depict the strain-crystallizing behavior of carbon black filled rubber, based on the experiments of \cite{rault2006}.
    Additional details on the extrapolation behavior of the PANN model are given in App.~\ref{sec:other_unfilled_NR}, along with a supplementary validation against experimental data from~\cite{candau2015}.
	All results presented in this contribution are obtained with PANN architectures, consisting of only a single hidden layer with 10 neurons in the convex and non-convex paths of the PICNNs.
    The loss weights are set to $w^{\te P} = 1$ and $w^\omega = 0.2$ for all calibrated models presented below, to account for the higher noise in the crystallinity measurements compared to the stress. 
    All calibrations were carried out on a local machine, equipped with a single Intel Core i7-1335U CPU. One training run took about $\SI{15}{\minute}$.
    
	\subsection{Unfilled natural rubber}
	
	For training the PANN, we rely on the experimental data of \cite{rault2006a} for NR under cyclic uniaxial loading at a constant stretch rate $|\dot{\lambda}|=11.67 \cdot 10^{-4}\,\SI{}{\per \second}$.
	Therein, stress and crystallinity data is available for hysteresis cycles of different stretch amplitudes, see Fig.~\ref{fig:Rault_6,5-5,0_omega_exp}. 
    
    \subsubsection{Calibration of the PANN}
    \label{sec:calibr_unfilled}
	The model calibration is carried out on the inner- and outermost hystereses, so that the remaining data can be used to evaluate the model on unseen loading paths.%
	\footnote{Naturally, the calibration may also be carried out considering only one single loading cycle. However, as shown in App.~\ref{sec:trainingdata_study}, including more than one loading and unloading sequence has been found to substantially enhance the  {generalization} of the model.}

	First, training is carried out considering both deformation invariants. The corresponding training results, as well as predictions for unseen load cycles, are shown in Fig.~\ref{fig:Rault_6,5-5,0_omega_exp_I1I2}.
	Second, in order to enable plausible predictions for multiaxial data, parameterization is repeated for an $\ici$-only model, cf. Remark~\ref{rem:i1only}, with the results given in Fig.~\ref{fig:Rault_6,5-5,0_omega_exp_I1}.
	As can be seen from Fig.~\ref{fig:Rault_6,5-5,0_omega_exp},
	both model versions exhibit very good agreement with the stretch--stress relations considered for calibration and satisfactory generalization behavior for the unknown load paths.
	Similarly, the courses of the DOC are in good qualitative agreement with the measurements for both model versions, although slender deviations are present.
	The performance of the model considering $\ici$ and $\iici$ is  marginally better compared to the $\ici$-only approach.
	However, this slightly better agreement in case of uniaxial data comes at the cost of unreliable predictions for multiaxial deformation states, see~\cite{dammass2025}. Therefore, the $\ici$-only version is considered in the remainder of this work, if not stated differently.
	
	\begin{figure}[h]
		\centering
		\subfloat[\centering $ \boldsymbol{\mathcal{I}} =(\ici, \iici^{3/2})$ as deformation input]{
			\label{fig:Rault_6,5-5,0_omega_exp_I1I2}
            \includegraphics[trim={0 1cm 0 0}, clip,scale=0.95]{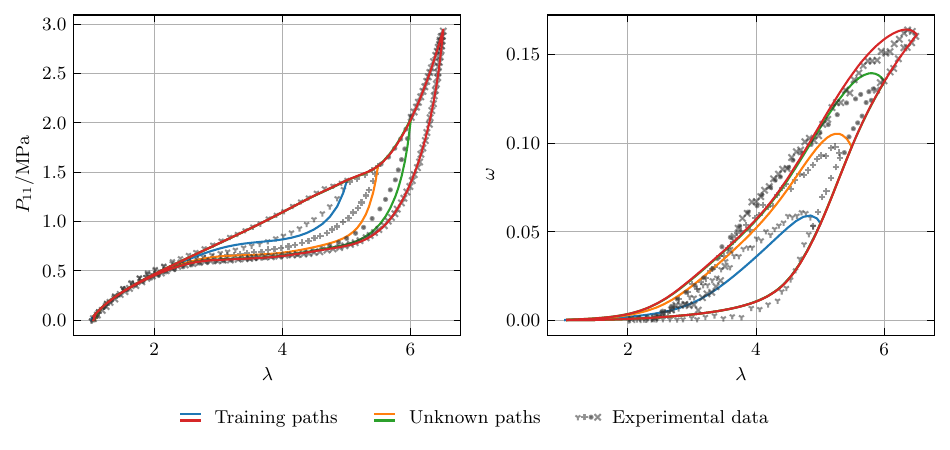}}
		\\
		\subfloat[\centering  $ \boldsymbol{\mathcal{I}} =\ici$ as deformation input]{
			\label{fig:Rault_6,5-5,0_omega_exp_I1}
			\includegraphics[scale=0.95]{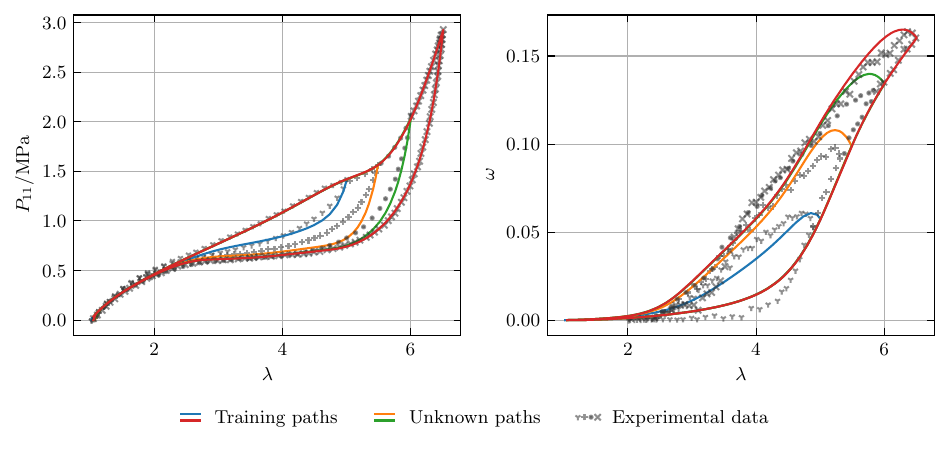}}
		\caption{PANNs trained on the experimental stress and crystallinity data for uniaxial tension of unfilled NR by~\cite{rault2006a}: \textbf{(a)} model version with $\ici$ and $\iici$ as deformation input and \textbf{(b)} only considering $\ici$.}
		\label{fig:Rault_6,5-5,0_omega_exp}
	\end{figure}
	
\begin{figure}
		\centering
		\includegraphics[trim={0 1.6cm 0 0}, clip,scale=0.95]{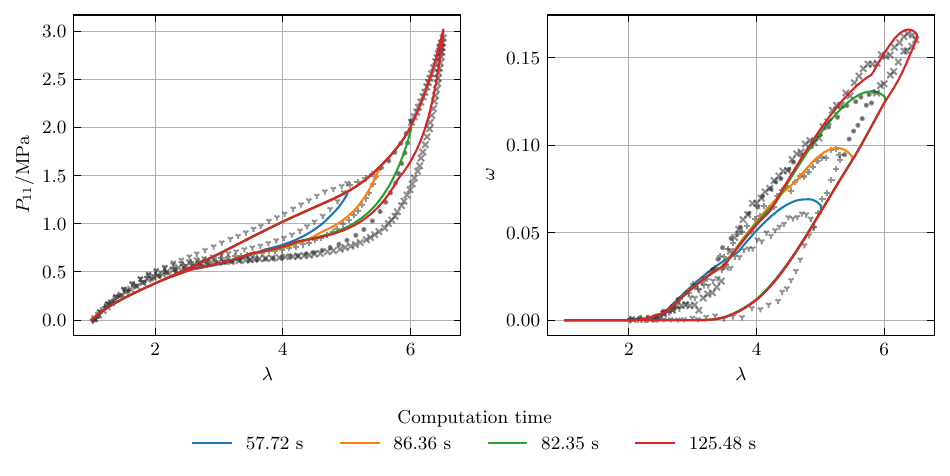}
		\caption[MAPC model for SIC considered as a benchmark]{MAPC model for SIC considered as a benchmark \cite{rastak2018}. 
			The model parameters are taken from the original contribution, calibrated on the exact considered set of experimental data by \cite{rault2006a}.
			Note that we use the symbols $\lambda, \omega$ to denote stretch and crystallinity on the continuum scale. In~\cite[Fig.~9]{rastak2018}, $\overline{\lambda}, \overline{\omega}$ are used for this purpose instead.
		}
        
\label{fig:MAPC}
\end{figure}

	To further put the performance of the developed model in a comparative context, we consider the maximal advance path constraint (MAPC)-based model for SIC, proposed by \cite{rastak2018}, as a statistical mechanics-based reference. 
	To the best of the author’s knowledge, this is among the most precise models for SIC in unfilled NR, incorporating rigorous micromechanical treatment of non-affine deformation between micro- and macro-scales.
	Therein, we use the discrete crystallinity representation with 37 quadrature points per hemisphere of the representative microsphere network, see \cite{bazant1986} for the considered quadrature scheme.
	The model parameters, taken from the original publication, were determined for this exact set of experimental data and integration scheme~\cite[p. 86]{rastak2018}.
	As shown in Fig.~\ref{fig:MAPC}, this model reflects the general traits of SIC, \ie, the hystereses of the stress and DOC over the loading cycles, with high quantitative agreement with the experimental data. 
{In direct comparison with the MAPC model, the proposed PANN model achieves an overall significantly better representation of the experimentally observed behavior.
In particular, the stress-stretch curves provide a substantially more accurate representation of the measurements, with virtually identical levels of precision for the crystallinity evolution. Quantitative performance data are provided in Table \ref{tab:MAE_r2_PANN_MAPC}. }
\begin{table}
 
    \centering
            \caption{   Performance quantification for MAPC-based model and PANN models considering both $\overline{I}_1$ and $\overline{I}_2$. Comparison against all hystereses in the data of \cite{rault2006a}.}

    \begin{tabular}{l | c c | c c} 
        & \multicolumn{2}{c|}{MAE} & \multicolumn{2}{c}{$r^2$} \\
        \hline
        & $P_{11}$ & $\omega$ & $P_{11}$ & $\omega$ \\
        \hline
        MAPC-based model \cite{rastak2018} & $0.09969$ MPa & $0.00648$ & $0.96207$  & $0.96243$\\
        PANN & $0.02496$ MPa & $ 0.00651$ & $0.99541$  & $0.96807$ \\
    \end{tabular}
    \label{tab:MAE_r2_PANN_MAPC}
\end{table}
Moreover, the PANN model turns out to be substantially more efficient in comparison with the considered benchmark model.
For the simulation of the material response under uniaxial tension, as presented in Figs.~\ref{fig:Rault_6,5-5,0_omega_exp}--\ref{fig:MAPC}, the computation time of the MAPC-based model was more than ten times larger than that of the PANN on the same machine.%
\footnote{To simulate the hysteresis loops shown in Figs.~\ref{fig:Rault_6,5-5,0_omega_exp}--\ref{fig:MAPC}, we implemented both the PANN and the MAPC-based model from~\cite{rastak2018} in a Python- and TensorFlow-based framework. 
For the latter, we closely followed the algorithmic strategies presented in~\cite{rastak2018}. 
To simulate one hysteresis loop, the computation times were in the order of 2~s for the PANN and in the order of 1~min for the MAPC-based model.
Nevertheless, we note note that there may still be opportunities to optimize the implementation of the latter model, which were beyond the scope of the present work.}

{ 
\subsubsection{Extrapolation study}
\label{sec:trainingdata_study}

For further insights into the performance, capabilities and limitations of the PANN model, we investigate the influence of the amount of data considered for calibration.
At the same time, we assess the model’s ability to extrapolate beyond deformation states encountered during training.
For this purpose, the stress and crystallinity measurements from~\cite{rault2006a} are considered, where only a single hysteresis loop is used for training in each case, while the others are employed as test paths.
The results of this study are presented in Fig.~\ref{fig:rault_study}.
Evidently, predictions for unseen loading paths are more reliable when those paths involve lower deformation levels than those encountered in training. 
However, we observe that with higher stretches encountered during training, the predictions for unseen deformation sequences become more accurate.

\begin{figure}
\centering
\includegraphics{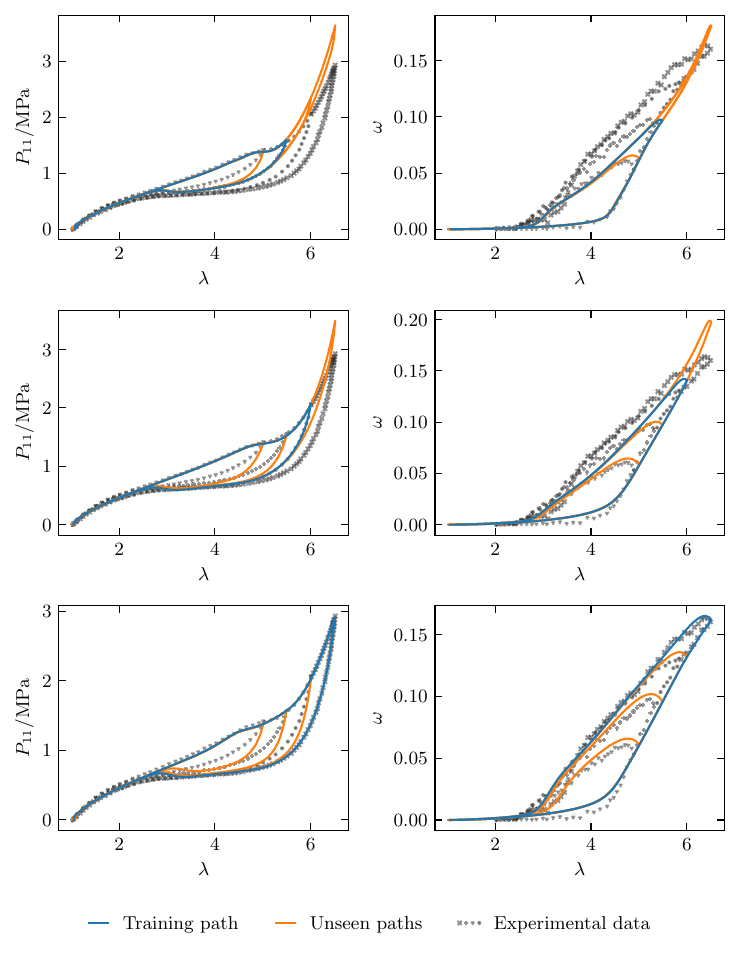}
\caption{  Predictions of the PANN trained on single loading and unloading paths of the experimental data from \cite{rault2006a}, whereas the remaining loading sequences are predicted. Here, the $\ici$-only formulation of the model is considered.
Owing to the physical constraints incorporated into the model, extrapolation beyond deformations encountered during training can yield predictions, that are, to a certain extent, within a reasonable range.
}
\label{fig:rault_study}
\end{figure}
}

\subsubsection{Finite element simulation}
\begin{figure}
\centering
\includegraphics{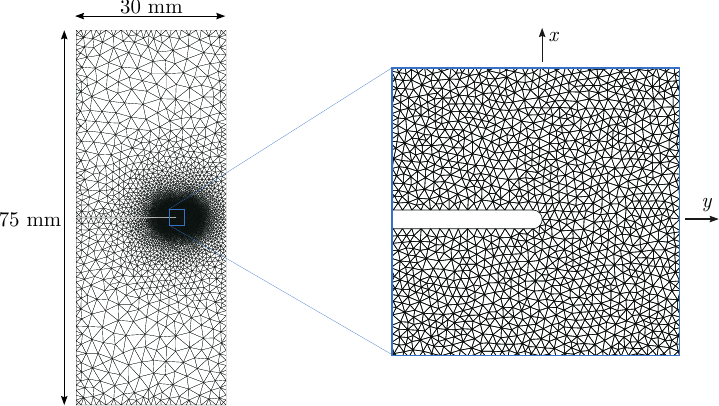}
\caption{Finite element mesh used for the SENT specimen with close-up of the notch region. The dimensions of the considered notch are $ \SI{20}{\milli \metre} \times \SI{0.2}{\milli \metre} $.}
\label{fig:SENT_mesh}
\end{figure}

\begin{figure}
\centering
\includegraphics[scale=0.9]{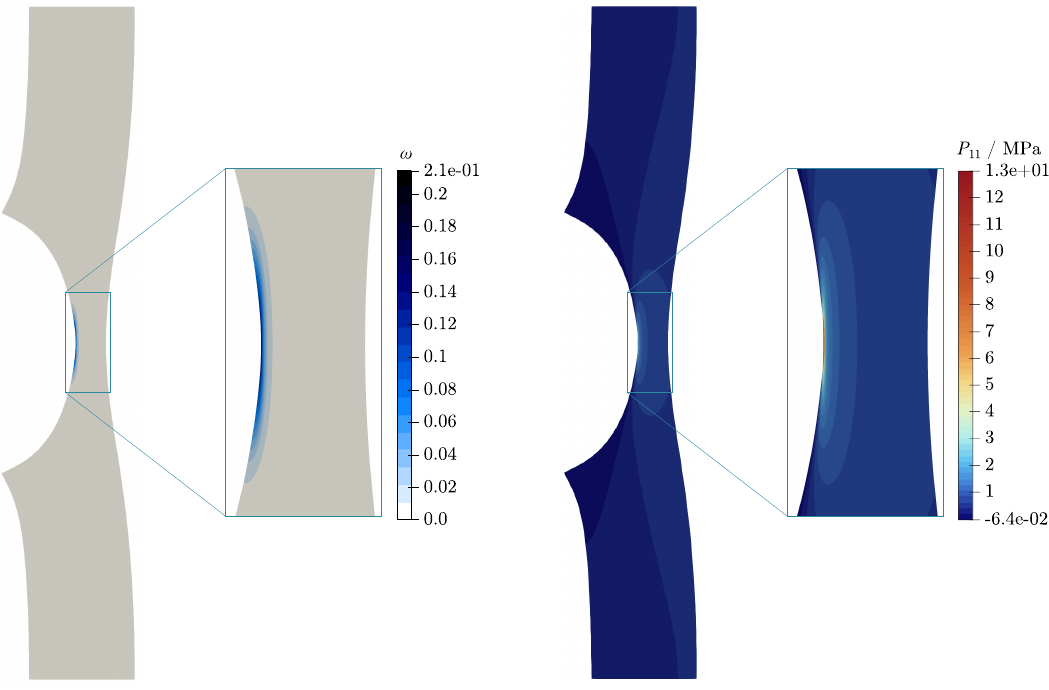}
\caption{DOC and stress distributions for a prescribed axial displacement of $\SI{75}{\milli \metre}$, depicted over the deformed specimen.}
\label{fig:SENT_results}
\end{figure}

\begin{figure}
\centering
\subfloat[\centering ]{
	\label{fig:DOC_notch_y_dir}
	\includegraphics[scale=0.95]{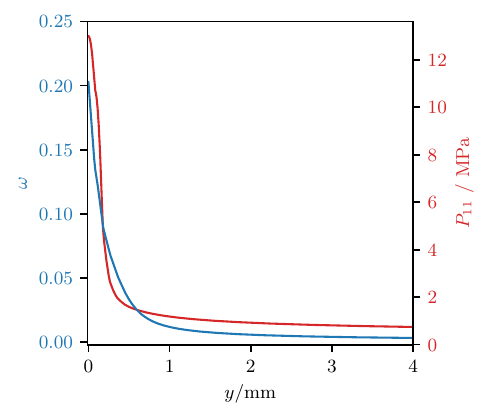}
}
\subfloat[\centering ]{
	\label{fig:DOC_notch_x_dir}
	\includegraphics[scale=0.95]{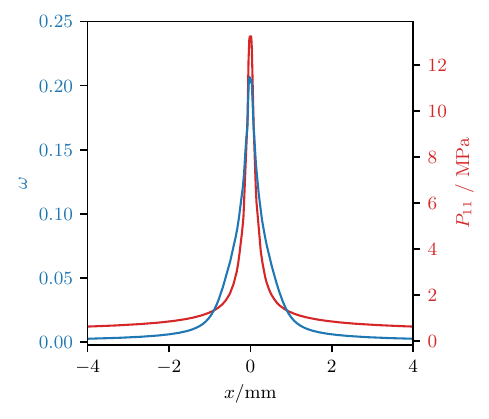}
}
\caption{DOC and stress distributions in the vicinity of the notch tip, plotted as functions of the coordinates in the configuration defined in Fig.~\ref{fig:SENT_mesh}. Evaluated for \textbf{(a)} y-direction, \ie, coaxial with the notch and \textbf{(b)} x-direction, \ie, perpendicular to the notch. A pronounced localization of both the stress and the DOC directly at the tip of the notch is observed.}
\label{fig:DOC_P_at_notch_SENT}
\end{figure}

To demonstrate the predictive capabilities of the model, we numerically investigate a single edge-notched specimen under tensile loading (SENT) as a representative benchmark for complex loading conditions, inspired by the experimental investigations from~\cite[Sec.~5.1]{xiang2022}.
To this end, we perform a finite element analysis~(FEA) employing the strategies introduced in Sec.~\ref{sec:FE_implementation}.
We consider a flat specimen of $(\SI{75}{\milli \metre} \times \SI{30}{\milli \metre}\times \SI{1}{\milli \metre})$, with a vertically centered notch $(\SI{20}{\milli \metre}\times\SI{0.2}{\milli \metre}$) with rounded tip. 
The geometry and the corresponding FE mesh, for which we consider bi-quadratic triangular elements, is shown in Fig.~\ref{fig:SENT_mesh}.
A vertical displacement load of $\SI{75}{\milli \metre}$ is uniformly applied at the top edge, while the bottom edge is fixed.
The numerical results for the DOC field and the stress component in tensile direction are shown in Fig.~\ref{fig:SENT_results} and Fig.~\ref{fig:DOC_P_at_notch_SENT}.
Overall, we observe reasonable spatial distributions of the field quantities, with strong localizations at the immediate vicinity of the notch tip. 
In particular, numerical predictions for the crystallinity field are in qualitative agreement with experimental observations of~\cite[Sec.~5.1]{xiang2022}.
Moreover, the magnitudes of the peak values are plausible, when compared to the MAPC-based model~\cite[Fig.~15]{rastak2018}, as well as evidence from similar experiments~\cite{rublon2014}.
The same applies for the spatial expansion of the crystalline domain.
Furthermore, the course of the DOC at the tip of the notch, see Fig~\ref{fig:DOC_P_at_notch_SENT}, is also found to be in qualitative agreement with both the predictions of existing models and experimental results, see~\cite[Fig.~15]{rastak2018}.

\subsection{Describing unfilled NR based on an automatically discovered pseudo-crystallinity variable}

Measuring crystallinity requires specialized, elaborate setups, and is therefore not always feasible.
Consequently, we investigate whether the state of a crystallizing rubber can also be described by an automatically discovered internal pseudo-crystallinity variable, which is not necessarily directly associated with the DOC.
We follow the strategy introduced above in Sec.~\ref{sec:training_withoutDOC}, and consider the scalar-valued pseudo-crystallinity variable $q$. We solely require non-negativity for $q$, which we enforce weakly during training, and employ the formulation of the loss function defined in \eqref{eq:Loss_negativity}.

The results of the training,
as well as predictions for unseen load sequences are shown in Fig.~\ref{fig:Rault_6,5-5,0_+omega_autogen}.
Similarly to Fig.~\ref{fig:Rault_6,5-5,0_omega_exp}, we again examine model formulations considering either both $\ici$ and $\iici$, or only $\ici$ as deformation input.
\begin{figure}[t]
\centering
\subfloat[\centering $ \boldsymbol{\mathcal{I}} =( \ici, \iici^{3/2} )$ as deformation input]{
	\includegraphics[trim={0 1cm 0 0}, clip]{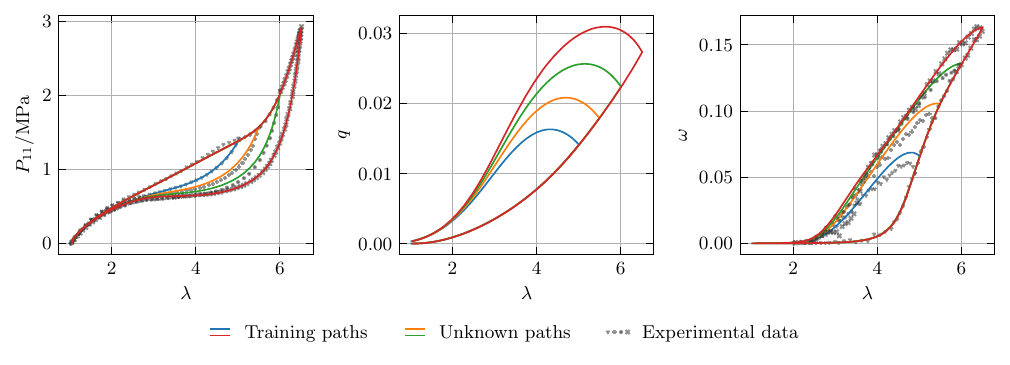}
	\label{fig:Rault_6,5-5,0_+omega_autogen_I1I2}
}\\ 
\subfloat[\centering $ \boldsymbol{\mathcal{I}}  = \ici$ as deformation input]{
	\includegraphics{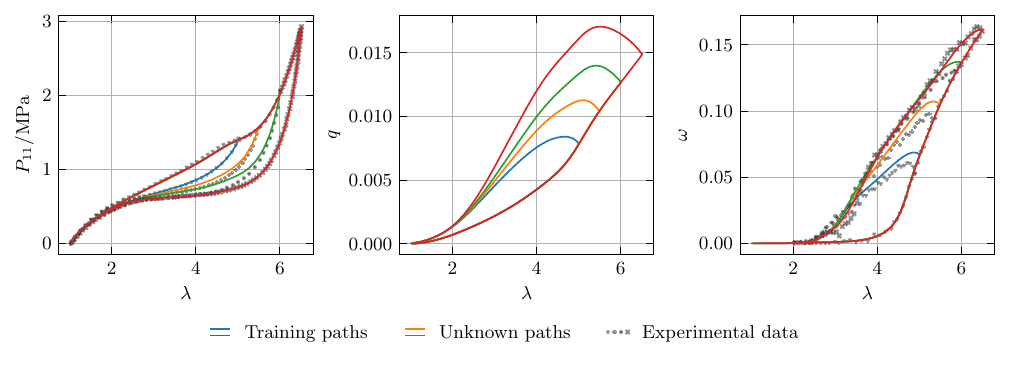}
	\label{fig:Rault_6,5-5,0_+omega_autogen_I1}
}
\caption{PANNs trained only on the experimental stress data for uniaxial tension of NR by \cite{rault2006a} with automatically discovered internal pseudo-crystallinity variable $q$, for which a mapping $\omega \approx \mathcal{Q}((\lambda_1, \lambda_2, \lambda_3), q)$ is found afterwards.}
\label{fig:Rault_6,5-5,0_+omega_autogen}
\end{figure}
Compared to the training based on stress and crystallinity data (Fig.~\ref{fig:Rault_6,5-5,0_omega_exp}), we obtain a slightly better agreement of both model versions with experimental stress data. Moreover, similarly to Fig.~\ref{fig:Rault_6,5-5,0_omega_exp}, the $\ici$ and $\iici$-variant achieves a slightly higher accuracy.
In both cases, the automatically discovered internal pseudo-crystallinity variable is one order of magnitude below the DOC. 
The evolution of the discovered internal variable $q$ only very roughly resembles the shape of the measured DOC hystereses.
Interestingly, however, the course of $q$ is non-monotonic during unloading, \ie, initially after reversing the deformation, the internal variable increases further. 
The same effect is also observed in experimentally-determined values of crystallinity and the model predictions shown in Fig.~\ref{fig:Rault_6,5-5,0_omega_exp}.

In a postprocessing step, we check whether the the pseudo-crystallinity $q$ can be linked to the actual DOC~$\omega$.
Specifically, we aim at finding a mapping 
\begin{equation}
\mathcal{Q}: \rs_{\geq 0} ^3 \times \mathbb{R}_{\geq 0} \longrightarrow [0, 1] \quad , \quad ((\lambda_1,\lambda_2,\lambda_3) ,  q)  \longmapsto\mathcal{Q}((\lambda_1,\lambda_2,\lambda_3), q) 
\comma
\end{equation}
such that $\mathcal{Q}((\lambda_1,\lambda_2,\lambda_3), q) \approx \omega$.
To identify a suitable $\mathcal{Q}$, we introduce an auxiliary feedforward neural network with internal normalization layers. Its parameters are identified minimizing the deviation between experimental crystallinity and pseudo-crystallinity, measured by the loss
\begin{equation}
    \mathcal{L^Q} = \frac{1}{N_\omega}\mathrm{MAE} \bigl(\mathcal{Q}(\pren\lambda_i,\pren q), \pren \omega^\mathrm{exp}\bigr) \commaf
\end{equation}
following a similar strategy as described for the PANN above in Sec.~\ref{sec:NN_training}.
Thereby, we ensure $\omega \in [0,1]$ by considering a sigmoid activation in the rearmost layer of $\mathcal{Q}$.
The posterior mappings found in Fig. \ref{fig:Rault_6,5-5,0_+omega_autogen} achieve {  adequate} approximation of the DOC.\par
 {We regard the possibility of calibrating the model without crystallinity measurements to be a valuable insight with potential practical applications. However, to account for SIC-related effects, such as increased fracture resistance, it is considered advantageous to employ a model parameterization calibrated to represent the DOC.}

\subsection{Filled natural rubber}
\label{sec:50phr}

\begin{figure}[h]
\centering
\includegraphics{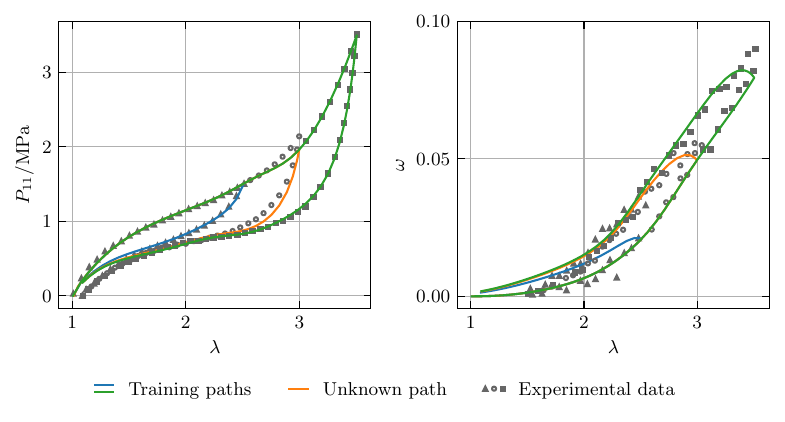}	
\caption{Calibrated PANN for the experimental data on filled NR containing $\SI{50}{\gram}$ carbon black per $\SI{100}{\gram}$ rubber by \cite{rault2006}. Here, both $\ici$ and $\iici$ are considered for the input of the network.
	Note that there is an inelastic permanent deformation at zero stress at the end of the loading sequences. This effect is not covered by the present model, causing slender deviations between experiment and model for $\lambda \lesssim 1.5$.}
\label{fig:Rault_50phr_I1I2}
\end{figure}

For technical applications, filled rubber holds significantly larger importance than its unfilled counterpart.
However, the presence of filler particles, with carbon black being the most prominent ones, significantly influences the microscopic behavior of the material, see e.g.,~\cite{kobayashi2015}.
As a consequence, classical, statistical mechanics-based models developed for unfilled systems are not suited for filled rubbers, and vice versa.
Conversely, the proposed PANN model is applicable for both material classes, as it considers only the effective macroscopic behavior, which is closely related for both material classes.
In what follows, the PANN model is calibrated based on experimental stress and crystallinity data from uniaxial tensile tests of filled NR containing $\SI{50}{\gram}$ carbon black per $\SI{100}{\gram}$ rubber, i.e., 50~phr, which are taken from~\cite{rault2006}.
The two loading cycles with highest and lowest maximal stretches are considered for training.
The remaining hysteresis serves to evaluate the model on a load path that has not been included in the calibration data, see Fig.~\ref{fig:Rault_50phr_I1I2}.

The model accurately captures the stress-stretch relation with high accuracy and provides sufficient precision for the testing path.
Furthermore, although a significant amount of noise seems to be present in the DOC measurements, the PANN achieves an appropriate evolution of the crystallinity. 

However, the model cannot completely capture the material response for $\lambda \lesssim 1.5$ for all hysteresis loops. 
This can be attributed to the presence of an inelastic permanent deformation at zero stress in the corresponding load sequences; this effect is not yet covered by the present model.

\section{Conclusion}
We present a thermodynamically consistent two potential modeling approach to strain-induced crystallization in rubber, based on an extension of the concept of generalized standard materials.
Equivalently, we provide a time-discrete variational framework, from which the governing equations of the mechanical problem can be derived in an alternative way. This formulation also serves as the basis for an FE implementation.
For the two potentials involved, we consider a physics-augmented neural network approach, incorporating important physical considerations. These are defined such that, most importantly, compliance with the second law of thermodynamics, objectivity and isotropy of the material response, as well as the fulfillment of normalization conditions for the material's initial state are guaranteed.
The resulting physics-augmented neural network model describes the effective constitutive behavior in a precise and efficient way, and generalizes between unfilled and filled rubber. 
The model calibration is carried out based on measurable deformation-stress-crystallinity data from uniaxial tension experiments by means of a Sobolev training.
The proposed model is further shown to  {generalize} well for deformation sequences, which were not included in the training, and is observed to simultaneously achieve higher efficiency and accuracy compared to established models found in literature.
In addition, we show that the proposed model can also  predict the stress response of a crystallizing rubber, when no experimental crystallinity data is available --~based on an automatically discovered internal pseudo-crystallinity variable, which can be linked to the crystallinity in a post-processing step.

In future work, the model could be extended to account for the thermal effects associated with strain-induced crystallinity evolution by means of a thermo-mechanically coupled formulation.
Moreover, complex fracture phenomena have been observed in NR, and an investigation could be pursued by combining the present model with fracture phase-field approaches, thereby extending our previous work \cite{dammass2025a}.

\section*{Supplementary material}
After publication of the final version of the manuscript, implementations of all the PANN models considered in this work, along with the specific parameters for the examples discussed in Secs.~\ref{sec:eval_exp_data} and \ref{sec:other_unfilled_NR}, will be available at \url{https://github.com/NEFM-TUDresden}.

\section*{Acknowledgments}
Support for this work was provided by the German Research Foundation (DFG) under grant KA 3309/20-1 (project 517438497).

\section*{CRediT author contribution statement}
\textit{Konrad Friedrichs:} Conceptualization, Formal analysis, Investigation, Methodology, Visualization, Software, Validation, Data Curation, Writing –- original draft, Writing –- editing.
\textit{Franz Dammaß:} Conceptualization, Formal analysis, Investigation, Methodology, Writing –- original draft, Writing –- editing, Supervision, Funding acquisition.
\textit{Karl A. Kalina:} Conceptualization, Formal analysis, Investigation, Methodology, Software, Writing –- original draft, Writing –- editing, Supervision.
\textit{Markus Kästner:} Resources, Writing –- editing, Supervision, Funding acquisition.

\section*{Declaration of competing interest}
There is no conflict of interest to declare.

\section*{Declaration of generative AI and AI-assisted technologies in the manuscript preparation process}
During the preparation of this work the authors used OpenAI’s language model ChatGPT to aid in text refinement and grammar checking. After using this tool, the authors reviewed and edited the content as needed and take full responsibility for the content of the published article.

\appendix
\section{Calibration based on another data set for unfilled rubber}
\label{sec:other_unfilled_NR}
\begin{figure}[h]
\centering
\includegraphics{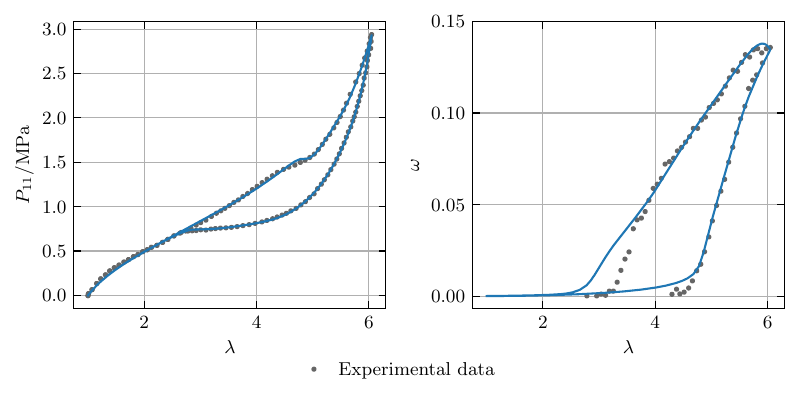}
\caption{Prediction of the PANN trained on the experimental data for unfilled NR \cite{candau2015}. Here, the deformation input is constituted of solely $\boldsymbol{\mathcal{I}}=\ici$. Note that, due to the lack of other loading cycles for this data set, the extrapolation capabilities of the PANN can not be evaluated here.}
\label{fig:Candau_NR}
\end{figure}

In addition to the measurements for unfilled NR presented in Sec.~\ref{sec:eval_exp_data}, the model is calibrated on the data from~\cite{candau2015}.
Fig. \ref{fig:Candau_NR} shows proper resemblance of the predicted stress hysteresis, and satisfactory matching of the crystallinity evolution. 

\bibliographystyle{abbrv}  
{\footnotesize \bibliography{bib_sicpann.bib}}	

\end{document}